\documentclass[11pt]{article}
\usepackage{arxiv}

\usepackage[utf8]{inputenc}
\usepackage[T1]{fontenc}
\usepackage[authoryear,round]{natbib}
\usepackage{textcomp}
\usepackage{amsmath,amssymb,amsfonts}
\usepackage{newtxmath}
\usepackage{graphicx}
\usepackage{booktabs}
\usepackage{array}
\usepackage{tabularx}
\usepackage{multirow}
\usepackage{here}
\usepackage{placeins}
\usepackage{bibunits}
\usepackage{microtype}
\usepackage{url}
\usepackage{xcolor}
\usepackage[hypertexnames=false,colorlinks=true,allcolors=blue]{hyperref}
\usepackage{doi}

\makeatletter
\renewcommand{\fnum@figure}{Fig.\nobreak\hspace{0.25em}\thefigure}
\makeatother

\fancypagestyle{supplementary}{%
  \fancyhf{}
  \fancyhead[C]{\footnotesize Supplementary Material of \textit{Evaluating AlphaEarth Foundations Embeddings for Wildfire Susceptibility Mapping}}
  \fancyfoot[L]{\small Zhuang et al., 2026: \textit{Supplementary Material}}
  \fancyfoot[R]{\small Page~\thepage{}~of~\SupplementLastPage}

}

\fancypagestyle{supplementaryfirst}{%
  \fancyhf{}
  \fancyhead[C]{\footnotesize Supplementary Material of \textit{Evaluating AlphaEarth Foundations Embeddings for Wildfire Susceptibility Mapping}}
  \fancyfoot[L]{\small Zhuang et al., 2026: \textit{Supplementary Material}}
  \fancyfoot[R]{\small Page~\thepage{}~of~\SupplementLastPage}

}
\title{Evaluating AlphaEarth Foundations Embeddings for Wildfire Susceptibility Mapping}

\author{%
\begin{minipage}{0.94\textwidth}
\centering
\normalfont\normalsize
\href{https://orcid.org/0009-0002-7338-7149}{\raisebox{-0.15ex}{\includegraphics[scale=0.06]{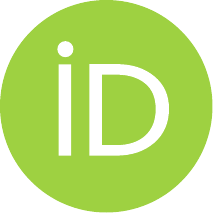}}}\hspace{0.3em}\textbf{Yuan Zhuang}\textsuperscript{1},\quad
\href{https://orcid.org/0000-0001-6825-3854}{\raisebox{-0.15ex}{\includegraphics[scale=0.06]{orcid.pdf}}}\hspace{0.3em}\textbf{Sanaa Hobeichi}\textsuperscript{2,3,4},\quad
\href{https://orcid.org/0000-0003-2789-3235}{\raisebox{-0.15ex}{\includegraphics[scale=0.06]{orcid.pdf}}}\hspace{0.3em}\textbf{Peng Shi}\textsuperscript{5},\quad
\href{https://orcid.org/0000-0002-6478-863X}{\raisebox{-0.15ex}{\includegraphics[scale=0.06]{orcid.pdf}}}\hspace{0.3em}\textbf{Fei Huang}\textsuperscript{1,*}
\\[0.8em]
\raggedright\normalfont\small
\textsuperscript{1}UNSW Sydney, School of Risk and Actuarial Studies, New South Wales, Australia\\
\textsuperscript{2}ARC Centre of Excellence for the Weather of the 21st Century, Australia\\
\textsuperscript{3}Climate Change Research Centre, University of New South Wales, Sydney, New South Wales, Australia\\
\textsuperscript{4}UNSW AI Institute, New South Wales, Australia\\
\textsuperscript{5}Department of Risk and Insurance, Wisconsin School of Business, University of Wisconsin--Madison, Madison, Wisconsin 53706, USA\\
\textsuperscript{*}Corresponding author: \href{mailto:feihuang@unsw.edu.au}{feihuang@unsw.edu.au}
\end{minipage}
}

\date{}
\renewcommand{\shorttitle}{Evaluating AlphaEarth Foundations Embeddings for Wildfire Susceptibility Mapping}

\hypersetup{
  pdftitle={Evaluating AlphaEarth Foundations Embeddings for Wildfire Susceptibility Mapping},
  pdfauthor={Yuan Zhuang, Sanaa Hobeichi, Peng Shi, Fei Huang},
  pdfkeywords={Wildfire susceptibility, AlphaEarth Foundations, spatial transferability, convolutional neural network, tabular foundation models}
}

\newcommand{\MainLastPage}{1}
\providecommand{\SupplementLastPage}{1}

\begin{document}
\maketitle
\def\floatpagepagefraction{1}
\def\textpagefraction{.001}

% NOTE (reviewer comment): this omits "Foundations", present in the full title below.
% Left as-is since shortening is plausibly deliberate (running-header space), but please confirm.

% Save full-page text width before \maketitle switches to two-column mode






% Abstract
\begin{abstract}
Wildfire susceptibility mapping typically relies on physical variables assembled from multiple remote-sensing, climate, and geospatial products. AlphaEarth Foundations (AEF) provides analysis-ready geospatial embeddings that may reduce this dependence on heavy harmonisation and task-specific feature engineering, but their value for wildfire susceptibility mapping has not been systematically evaluated. Using Victoria, Australia (2017--2025), as a case study, we show that AEF embeddings can reconstruct commonly used variables in wildfire susceptibility analysis with high accuracy. In downstream susceptibility models trained on satellite-derived fire occurrence data, embedding-based susceptibility models achieve ROC-AUC values above 0.92 and consistently identify high wildfire susceptibility across eastern Victoria, particularly Gippsland and the north-eastern uplands, with additional localized hotspots in central and northwestern Victoria. A key feature of AEF embeddings is their strong near-region transferability within climatically similar regions. When embedding-based models trained in Victoria are applied to Canberra and Western Sydney--Blue Mountains, ROC-AUC improves by around 4\% at Canberra and declines by around 2\% at Western Sydney--Blue Mountains, compared with a mean decrease of approximately 25\% for physical-variable models. These findings provide practical guidance for using AEF embeddings and lay a foundation for scalable wildfire susceptibility mapping workflows for downstream users such as government agencies and (re)insurers.
\end{abstract}


% Keywords

\newcommand{\TP}{\mathrm{TP}}
\newcommand{\FP}{\mathrm{FP}}
\newcommand{\TN}{\mathrm{TN}}
\newcommand{\FN}{\mathrm{FN}}

% Main text
\section{Introduction}\label{sec:intro}
Extreme wildfires are increasing in frequency and severity globally \citep{cunninghamIncreasingFrequencyIntensity2024,richardson2022global}. These trends place growing pressure on ecosystems \citep{bowmanFireEarthSystem2009}, communities \citep{filkovImpactAustraliasCatastrophic2020,grantLongtermHealthEffects2022}, and insurance markets \citep{Sigma_1_2026}. Therefore, identifying which locations are inherently susceptible to 
wildfire has become an increasingly central concern for governments, emergency planners, and (re)insurers.

In wildfire risk analysis, susceptibility mapping provides a spatially explicit framework for identifying landscapes that are more prone to fire occurrence. It is typically based on the premise that areas where wildfires have occurred in the past, together with areas sharing similar environmental and physical characteristics, provide useful indications of where fire events may be more likely to occur \citep{leuenbergerWildfireSusceptibilityMapping2018}. The resulting maps usually assign each pixel or spatial unit a relative susceptibility score. Such scores can help individuals, insurers, and public agencies understand spatial exposure to wildfire and support targeted mitigation policies.

Over the past two decades, the proliferation of remote sensing data and Geographic Information Systems (GIS) has enabled machine learning to become a major branch of wildfire susceptibility mapping. These approaches typically assemble fire occurrence records and construct predictor variables that describe environmental conditions conducive to ignition \citep{ejazComprehensiveSurveyMachine2025}. 
A model is then trained to learn a mapping from the covariates to observed fire occurrences and applied across all pixels or patches in the study region to produce a continuous susceptibility surface \citep{jainReviewMachineLearning2020}. Existing studies have used a wide range of machine learning algorithms, including MaxEnt \citep{moritzClimateChangeDisruptions2012}, support vector machines \citep{benzougaghHybridSVMXGBoost2026}, tree-based ensemble methods \citep{haydarMappingClimateChange2025,luuIntegratingSusceptibilityMaps2024,zeroualiAdvancedForestFire2025}, and shallow neural networks \citep{singhaIntegratingGeospatialRemote2024,tienbuiHybridArtificialIntelligence2017}, with more recent work increasingly adopting deep learning approaches \citep{xuDeepLearningWildfire2025} such as convolutional neural networks (CNNs) \citep{bjanesDeepLearningEnsemble2021,hakimDualstageWildfireRisk2025,jiangWildfireRiskAssessment2024,zhangForestFireSusceptibility2019,zhangDeepNeuralNetworks2021} that leverage spatially structured inputs to capture neighbourhood-level patterns.

Despite these methodological advances, assembling and harmonising suitable prediction datasets remains a major practical bottleneck in wildfire susceptibility mapping \citep{jainReviewMachineLearning2020,chicasWhoAreActors2022,ejazComprehensiveSurveyMachine2025}. Most current data-driven approaches rely on physical variables derived from Earth observation products, climate reanalysis, and other geospatial sources \citep{ejazComprehensiveSurveyMachine2025}. Although many of these products are publicly available, their use in susceptibility mapping is not trivial. First, predictor selection requires both scientific judgement about which variables are relevant to fire occurrence and whether those variables are available at appropriate spatial resolutions. Even commonly used predictors may be too coarse, derived from discontinued products, or restricted to particular regions. Second, predictors are often produced by different sensors or systems, and some require additional processing or derivation before they can be used as modelling inputs \citep{chicasWhoAreActors2022}. As a result, substantial harmonisation and feature engineering are typically required before any modelling can proceed \citep{maHarvestingAlphaEarthBenchmarking2026}. Finally, susceptibility models may not transfer reliably beyond the regions in which they were developed \citep{bekarCrossregionalModellingFire2020}, so applying them elsewhere may require a tailored data pipeline and a newly trained model. This repeated redevelopment raises the entry barrier, particularly for stakeholders who could benefit from spatial risk information but lack the necessary technical capacity.

Earth foundation models have recently emerged as a promising source of alternative geospatial feature representations that may help address these data-related barriers. Trained on large-scale Earth observation archives, these models learn rich representations of the physical landscape \citep{zhuFoundationsEarthFoundation2026}, and the resulting embeddings can then be extracted as off-the-shelf features for a wide range of downstream tasks \citep{yeAnyModelAny2026}. Among existing embedding products, AlphaEarth Foundations (AEF) is a publicly available and analysis-ready example \citep{brownAlphaEarthFoundationsEmbedding2025}. AEF was trained using over 3 billion observations from nine gridded data sources and two unstructured text sources, including optical imagery, climate variables, topography, and land cover information. The resulting product is distributed through Google Earth Engine as annual embedding layers for 2017--2025, covering terrestrial land surfaces and shallow waters globally. Each annual layer provides 64-dimensional vectors at 10~m resolution that summarise multi-sensor information for the corresponding year.

Since its release, AEF embeddings have been applied across a range of domains, including agriculture, socio-economics, environmental monitoring, ecology, and hydrology \citep{houAlphaEarthFoundationsAEF2026}. Because the training sources of AEF closely overlap with those commonly used to derive wildfire susceptibility predictors, the resulting embeddings may provide off-the-shelf features for downstream fire occurrence models. However, it remains unclear whether AEF embeddings encode the environmental conditions associated with wildfire susceptibility, how such information should be extracted by downstream susceptibility models, and whether the resulting models can transfer beyond the original study region. This study therefore proposes three research questions:
\begin{enumerate}
	\item To what extent do general-purpose AEF embeddings encode wildfire-relevant environmental information?
	\item How do different downstream modelling strategies compare when trained on AEF embeddings and satellite-derived fire occurrence data for wildfire susceptibility mapping?
	\item To what extent do AEF-based susceptibility models transfer to new regions, and how does their transferability compare with models trained on conventional physical variables?
\end{enumerate}

This study addresses these questions using Victoria, Australia, as a case study. We first evaluated AEF embeddings under a time-aggregated susceptibility modelling setting, using both cell-level tabular models and neighbourhood-based CNNs. By comparing their predictive performance with that of physical variables and conducting a reconstruction analysis of traditional predictors from the embeddings, we examined both what wildfire-relevant information is encoded in the embeddings and whether this information is sufficient for susceptibility mapping. We then introduced sequence-informed modelling strategies using annual feature layers to assess whether interannual information could further improve the use of AEF embeddings. Finally, we evaluated transferability by applying models trained in Victoria to multiple regions across Australia, including areas from different climate zones. Collectively, our analyses provide the first systematic evaluation of Earth foundation model embeddings for wildfire susceptibility mapping and demonstrate their potential value as reusable geospatial data assets for supporting more scalable wildfire risk mapping workflows in hazard assessment, exposure mapping, and climate adaptation planning.

The remainder of this paper is structured as follows. Section~\ref{sec:study_area} describes the study area and the data used in this study. Section~\ref{sec:methods} presents the modelling settings and evaluation framework. Section~\ref{sec:results} reports the susceptibility modelling, reconstruction, and transferability results. Section~\ref{sec:discussion} discusses the main findings and their limitations, and Section~\ref{sec:conclusion} concludes the paper.

\section{Data}\label{sec:study_area}
\subsection{Study area and Transfer Sites}
\label{sec:study_area_transfer}

\begin{figure}[htbp]
	\centering
	\includegraphics[width=\textwidth]{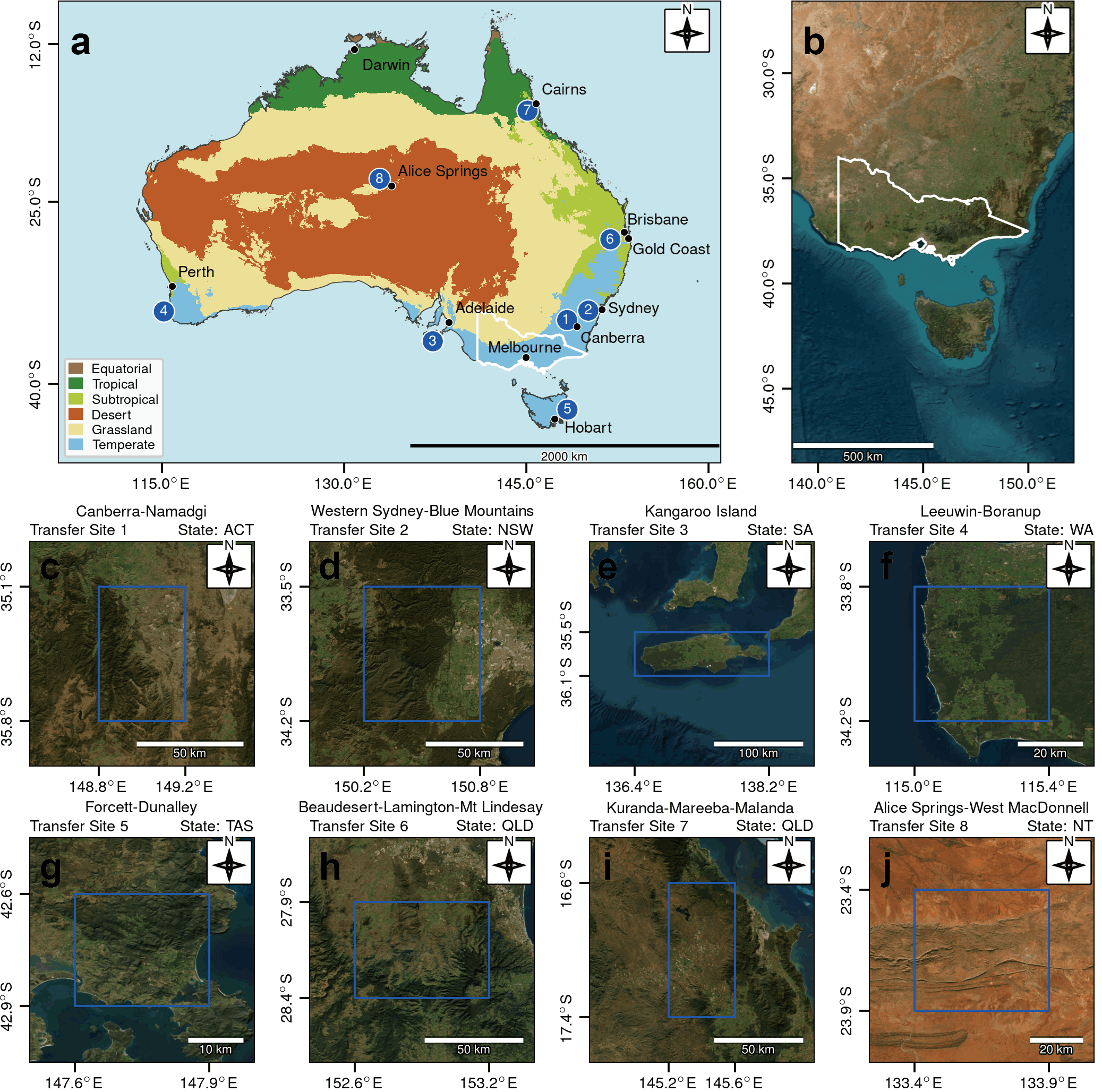}
	\caption{Study area (Victoria, Australia) and transfer sites across diverse climate zones. Climate zone classification sourced from \citet{bom2024climateclassification}, based on the 1991--2020 climatological reference period.}
	\label{fig:study_area}
\end{figure}

Victoria is located in southeastern Australia (34\textdegree{}--39\textdegree{}S, 141\textdegree{}--150\textdegree{}E) and covers approximately 227,000~km\textsuperscript{2} (Fig.~\ref{fig:study_area}a, b). The state spans a wide environmental gradient, from the highly urbanised Melbourne region to the sparsely populated alpine and forested landscapes of the Great Dividing Range and the Victorian Alps. Its climate also varies substantially. The southern coastal region is broadly temperate, with cool, wet winters and warm, dry summers, while northwestern Victoria is drier and semi-arid. Together, this landscape and climatic diversity provides a suitable setting for evaluating whether embedding-based susceptibility models perform consistently across heterogeneous environments within a single source region. Victoria is also historically fire-prone, with severe bushfire events including Ash Wednesday in 1983, Black Saturday in 2009, the 2019--2020 Black Summer, and the January 2026 bushfires \citep{ffmvic2021pastbushfires,vicgov2026bushfires}. Their repeated impacts, including extensive burned areas, loss of life, and major damage to settlements and ecosystems, have motivated recent research on wildfire susceptibility mapping in the region \citep{abdollahiExplainableArtificialIntelligence2023,hosseiniGeneExpressionProgramming2021,zhengMappingBushfireRisk2025}. These characteristics make Victoria a suitable source domain for developing and evaluating the wildfire susceptibility models in this study.

A further consideration is whether susceptibility models trained in one region can transfer reliably to ecologically distinct environments. Beyond Victoria, eight transfer sites were selected across Australia (Fig.~\ref{fig:study_area}c--j). The sites were arranged along a gradient of increasing geographic and climatic distance from the source domain: beginning with climatically similar and proximate locations in Canberra and Western Sydney, progressing to more distant but still temperate settings in South Australia, Tasmania, and Western Australia, and extending to climatically distinct out-of-domain environments in subtropical-to-tropical Queensland and the semi-arid interior of the Northern Territory. This gradient design enables a systematic characterisation of the transferability limits of AEF embeddings in wildfire susceptibility mapping.

\subsection{Physical Wildfire Driving Factors} \label{subsec:hand}
To provide a task-specific reference for evaluating the wildfire-relevant information encoded in AEF embeddings, we constructed a comprehensive set of physical driving factors commonly used in wildfire susceptibility mapping literature \citep{marjaniThreeDecadesGeoAI2026,xuDeepLearningWildfire2025}. The initial variable pool included 35 predictors across ten thematic categories (Table~\ref{tab:variables}). All predictors were harmonised to a common 1~km reference grid under Australian Albers (EPSG:9473). Detailed variable quantification and visualization are provided in Supplementary Material~\ref{sec:app_variables}.

From a wildfire-susceptibility perspective, these factors characterise complementary dimensions of fire-conducive conditions. Topographic variables describe terrain structure and exposure, which can influence local fuel conditions and potential fire spread. Land-cover classes, vegetation indices, atmospheric variables, and soil-moisture variables jointly describe fuel type, fuel availability, and moisture stress. Temperature, precipitation, wind speed, and fire-weather indices capture meteorological conditions related to fuel dryness, ignition likelihood, and potential fire behaviour. Proximity variables approximate human accessibility and spatial separation from water bodies. In addition, five variables are marked as fire-season summaries (F.S.), which are designed to capture extreme conditions using the counts of high-risk days.

These variables were organised into two datasets to support different modelling objectives. 
First, for each grid cell $i$ and calendar year $t \in \{2017, \dots, 2025\}$, 
we constructed a \textit{yearly dataset} 
$\mathbf{X}^{\mathrm{yr}} = \{\mathbf{x}_{i,t}\}$. 
Here, $\mathbf{x}_{i,t} \in \mathbb{R}^{V}$ denotes the physical feature vector 
for grid cell $i$ in year $t$, where $V$ is the number of physical variables and 
$x_{i,t}^{(v)}$ denotes the value of variable $v$. In compiling the yearly dataset, dynamic variables were summarised at the annual scale (e.g.\ annual mean temperature $\overline{T}_{i,t}$, annual total precipitation $P_{i,t}$), while 
static variables remain unchanged across years. We then derived a \textit{cross-sectional dataset} $\mathbf{X}^{\text{cs}} = \{\mathbf{x}_i\}$ by aggregating the yearly variables over the full study period or, where appropriate, by applying custom cross-year computation as detailed in Supplementary Material~\ref{sec:app_variables}. This produced a single feature vector for each grid cell, following the standard cross-sectional structure commonly used in wildfire susceptibility mapping \citep{jainReviewMachineLearning2020}. 
In the analyses later in this paper, $\mathbf{X}^{\text{yr}}$ was used for reconstruction study (Section \ref{sec:reconstruction}) and sequence-informed modelling (Section \ref{sec:temporal_methods}), whereas $\mathbf{X}^{\text{cs}}$ provided the main input for time-aggregated susceptibility modelling (Sections \ref{sec:cs_methods} and \ref{sec:victoriaevaluation}) and transferability evaluation (Section \ref{sec:transfer_methods}).

To address multicollinearity among the initial candidate features, variance inflation factor (VIF) screening was applied to $\mathbf{X}^{\text{cs}}$. Predictors with VIF exceeding 10 \citep{xuDeepLearningWildfire2025} were iteratively removed, yielding a final set of 22 retained predictors for both datasets, as indicated in Table~\ref{tab:variables}.

\begin{table}[htbp]
	\centering
	\small
	\caption{Candidate physical wildfire driving factors, with data sources, native resolutions, and VIF screening outcomes. Variables with VIF $<10$ were retained.}\label{tab:variables}
	\setlength{\tabcolsep}{4pt}
	\begin{tabularx}{\textwidth}{@{} >{\raggedright\arraybackslash}p{1.6cm} >{\raggedright\arraybackslash}p{4.6cm} l >{\raggedright\arraybackslash}X >{\raggedright\arraybackslash}p{1.3cm} c @{}}
		\toprule
		\textbf{Category} & \textbf{Variable} & \textbf{Code} & \textbf{Source} & \textbf{\shortstack{Resolution}} & \textbf{\shortstack{Retained}} \\
		\midrule
		\multirow[c]{4}{1.8cm}{Topography}
		& Elevation   & ELEV  & \multirow[c]{4}{*}{\parbox{\linewidth}{Australia 3 arc-sec SRTM DEM \\ \citep{gallant2009srtmdem}}} & 3 arc-sec & N \\
		& Slope       & SLOPE  &  & 3 arc-sec & Y \\
		& Northness   & NORTH &  & 3 arc-sec & Y \\
		& Eastness    & EAST  &  & 3 arc-sec & Y \\
		\midrule
		\multirow[c]{5}{1.8cm}{Land cover}
		& Cropland  & LC\_CRP & \multirow[c]{5}{*}{\parbox{\linewidth}{MODIS MCD12Q1 \\ \citep{friedl2022mcd12q1}}} & 500~m & Y \\
		& Forest    & LC\_FRS &  & 500~m & N \\
		& Grassland & LC\_GRS &  & 500~m & Y \\
		& Savanna   & LC\_SVN &  & 500~m & Y \\
		& Shrubland & LC\_SHR &  & 500~m & Y \\
		\midrule
		\multirow[c]{3}{1.8cm}{Precipitation}
		& Total annual precipitation      & PREC  & \multirow[c]{3}{*}{\parbox{\linewidth}{SILO \citep{jeffrey2001silo}}} & $0.05^\circ$ & Y \\
		& Max consecutive dry days (F.S.) & CDD   &  & $0.05^\circ$ & Y \\
		& Precipitation seasonality       & PRECS &  & $0.05^\circ$ & Y \\
		\midrule
		\multirow[c]{8}{1.8cm}{Temperature and radiation}
		& Mean temperature                & TMEAN & \multirow[c]{8}{*}{\parbox{\linewidth}{SILO \citep{jeffrey2001silo}}} & $0.05^\circ$ & Y \\
		& Max temperature                 & TMAX  &  & $0.05^\circ$ & N \\
		& Mean monthly temperature range  & MTR   &  & $0.05^\circ$ & N \\
		& Temperature annual range        & TAR   &  & $0.05^\circ$ & Y \\
		& Temperature seasonality         & TS    &  & $0.05^\circ$ & N \\
		& Max consecutive hot days (F.S.) & CHD   &  & $0.05^\circ$ & Y \\
		& Isothermality                   & ISO   &  & $0.05^\circ$ & N \\
		& Mean solar radiation            & SR  &  & $0.05^\circ$ & Y \\
		\midrule
		Wind
		& Mean wind speed & WIND & \parbox{\linewidth}{TerraClimate \citep{abatzoglou2018terraclimate}} & $(1/24)^\circ$ & Y \\
		\midrule
		\multirow[c]{4}{1.8cm}{Vegetation}
		& Mean NDVI & NDVI    & \multirow[c]{2}{*}{\parbox{\linewidth}{MODIS MOD13Q1 \citep{didan2021mod13q1}}}   & 250~m & N \\
		& Max NDVI  & NDVIMAX &  & 250~m & Y \\
		& Mean LAI  & LAI     & \multirow[c]{2}{*}{\parbox{\linewidth}{VIIRS VNP15A2H \citep{myneni2023vnp15a2h}}} & 500~m & N \\
		& Max LAI   & LAIMAX  &  & 500~m & N \\
		\midrule
		\multirow[c]{3}{1.8cm}{Fire weather}
		& Mean FFDI                            & FFDI   & \multirow[c]{3}{*}{\parbox{\linewidth}{\citet{copernicus2019firedanger}}} & $0.25^\circ$ & N \\
		& Max consecutive FFDI $>$ P90 (F.S.) & CFFDI  &  & $0.25^\circ$ & Y \\
		& Days FFDI $>$ 50 (F.S.)             & FFDI50 &  & $0.25^\circ$ & Y \\
		\midrule
		\multirow[c]{2}{1.8cm}{Atmospheric demand}
		& Mean VPD deficit            & VPD & \multirow[c]{2}{*}{\parbox{\linewidth}{SILO \citep{jeffrey2001silo}}} & $0.05^\circ$ & N \\
		& Mean water deficit (Morton) & WD  &  & $0.05^\circ$ & N \\
		\midrule
		\multirow[c]{3}{1.8cm}{Soil moisture}
		& Mean soil moisture                   & SM    & \multirow[c]{3}{*}{\parbox{\linewidth}{\citet{bom2024awo}}} & $0.05^\circ$ & N \\
		& Minimum soil moisture                & SMMIN &  & $0.05^\circ$ & Y \\
		& Max consecutive dry soil days (F.S.) & CDSD  &  & $0.05^\circ$ & Y \\
		\midrule
		\multirow[c]{2}{1.8cm}{Proximity}
		& Distance to major road  & DROAD  & \citet{osm2024}          & --- & Y \\
		& Distance to waterbody   & DWATER & \citet{bom2024geofabric} & --- & Y \\
		\bottomrule
		\multicolumn{6}{@{}p{\textwidth}@{}}{\footnotesize Note: F.S. indicates fire-season summaries, computed for the Australian fire season (Jan--Apr and Oct--Dec). Y = retained after VIF screening; N = removed.}
	\end{tabularx}
\end{table}

\subsection{AlphaEarth Foundations (AEF) Embeddings}

AlphaEarth Foundations (AEF) embeddings \citep{brownAlphaEarthFoundationsEmbedding2025}
are a geospatial data product developed by Google and Google DeepMind. The dataset provides annual 64-dimensional embedding vectors at a native 10~m resolution, encoding multi-spectral, multi-modal Earth observation information for a full calendar year. The collection is accessible as an image collection in Google Earth Engine (GEE) and as Cloud-Optimised GeoTIFFs (COGs) via a public Google Cloud Storage (GCS) bucket \citep{gee2026aef}.

For storage efficiency, embedding values are provided as signed 8-bit integer vectors $\mathbf{q}_{j,t} \in \{-128, \dots, 127\}^{64}$, where $j$ indexes each 10~m pixel and $t \in \{2017, \dots, 2025\}$. To recover the original embedding values, raw \texttt{int8} tiles covering Victoria were de-quantised to $[-1, 1]^{64}$ via
\begin{equation}
	\tilde{e}_{j,t,k}
	= \left(\frac{q_{j,t,k}}{127.5}\right)^{2}
	\cdot \operatorname{sign}(q_{j,t,k}), \quad k = 1, \dots, 64,
\end{equation}
where $q_{j,t,k}$ denotes the $k$-th element of $\mathbf{q}_{j,t}$. The de-quantised 10~m vectors $\tilde{\mathbf{e}}_{j,t} \in \mathbb{R}^{64}$ were then aggregated to the 1~km reference grid defined in Section \ref{subsec:hand}, by summing over all contributing 10~m pixels within each 1~km grid cell and L2-normalising the resulting vector, following the pyramid downsampling convention \citep{google2026aef}:
\begin{equation}
	\mathbf{e}_{i,t}
	= \frac{\displaystyle\sum_{j \in \mathcal{N}(i)} \tilde{\mathbf{e}}_{j,t}}
	{\left\|\displaystyle\sum_{j \in \mathcal{N}(i)}
		\tilde{\mathbf{e}}_{j,t}\right\|_{2}},
\end{equation}
where $\mathcal{N}(i)$ denotes the set of 10~m pixels falling within 1~km grid $i$.

Beyond the yearly dataset $\mathbf{E}^{\text{yr}} = \{\mathbf{e}_{i,t}\}$, a cross-sectional embedding dataset $\mathbf{E}^{\text{cs}} = \{\mathbf{e}_{i}\}$ was derived by averaging each grid cell's annual embedding vectors across all
available years:
\begin{equation}
	\mathbf{e}_{i}
	= \frac{1}{T}\sum_{t=1}^{T} \mathbf{e}_{i,t},
\end{equation}
where $T$ denotes the number of years. In both datasets, all 64 embedding dimensions were retained as features. $\mathbf{E}^{\text{yr}}$ and $\mathbf{E}^{\text{cs}}$ served as embedding-based counterparts to $\mathbf{X}^{\text{yr}}$ and $\mathbf{X}^{\text{cs}}$, respectively, and were used in exactly the same analytical roles throughout this paper.

\subsection{Wildfire Inventory and Sampling Strategy} \label{subsec:fire}

Fire occurrence records for Victoria were obtained from the MODIS Active Fire product (MCD14ML) distributed by NASA's Fire Information for Resource Management System (FIRMS) \citep{giglio2016collection}. The retrieved dataset consists of point locations (hotspots) detected by the MODIS sensors aboard the Terra and Aqua satellites over the period 2017--2025. Each hotspot record includes geographic coordinates, acquisition date, and confidence level (0--100\%). To reduce commission errors, only detections with confidence $\geq 80\%$ were retained. The filtered point records were rasterised to the 1~km reference grid by assigning a value of 1 to any grid cell containing at least one qualified hotspot during 2017--2025. This yielded 18,140 fire-occurrence cells, corresponding to approximately 8\% of the 1 km reference grid cells covering Victoria.

An equal number of non-fire cells were selected as pseudo-absence samples and assigned a label of 0. The choice of pseudo-absence samples is important because poorly selected non-fire cells can make it harder for the model to distinguish fire-prone from less fire-prone areas. Two spatial constraints are commonly imposed when drawing negative samples in wildfire susceptibility mapping. First, a minimum separation distance between positive and negative samples prevents the model from receiving contradictory supervision signals at environmentally similar locations \citep{hakimDualstageWildfireRisk2025,jiangWildfireRiskAssessment2024}. A threshold of 3~km was enforced here, consistent with empirical evidence in the literature \citep{heApplicationBurnedArea2026} and also supported by the spatial autocorrelation structure of AEF embeddings characterised in the semivariogram analysis (Supplementary Material~\ref{app:spatial_autocorrelation}). Second, spatial dispersion among pseudo-absence samples reduces excessive clustering and improves their coverage of the study area \citep{heApplicationBurnedArea2026,jiangWildfireRiskAssessment2024}. Eligible pseudo-absence cells were sampled using an auxiliary 3~km uniform grid, with at most one 1~km candidate cell selected from each auxiliary cell. The resulting sample was then randomly trimmed to match the number of fire-occurrence cells.

\section{Methods}\label{sec:methods}

\subsection{Overview of Victoria Experimental Design} \label{sec:overview}

The labelled dataset was partitioned into a training set (70\%) and a held-out test set (30\%) using a stratified random split, preserving the 1:1 positive-to-negative ratio in both subsets. This procedure was repeated across 10 independent splits to obtain stable performance estimates. All reported metrics represent the mean test-set performance across these runs, with variability reported in Supplementary Material~\ref{app:additional_results}. For all tunable models, hyperparameters were tuned on the training set using Optuna \citep{akibaOptunaNextgenerationHyperparameter2019} with a Tree-structured Parzen Estimator (TPE) sampler and 5-fold cross-validation, where the mean ROC-AUC across validation folds served as the optimisation objective. As the within-Victoria evaluation targets interpolation over a spatially exhaustive grid, random stratified splits are appropriate for map accuracy assessment \citep{wadouxSpatialCrossvalidationNot2021}. Spatial generalisation is evaluated through the transferability analysis in Section~\ref{sec:transfer_methods}.

\subsection{Time-Aggregated Susceptibility Modelling}\label{sec:cs_methods}

The first modelling setting considered time-aggregated susceptibility prediction. For physical variables and AEF embeddings, we use cross-sectional datasets $\mathbf{X}^{\mathrm{cs}}$ and $\mathbf{E}^{\mathrm{cs}}$ here. Across all models in this time-aggregated setting, the physical representation contained 22 input features, whereas the AEF representation contained 64 embedding dimensions. These representations were evaluated under two complementary input structures: cell-level models, which treated each grid cell as an independent observation, and spatial neighbourhood models, which incorporated surrounding grid cells through local image patches.

\paragraph{Cell-level models} Three tabular model families were evaluated to compare these feature representations across different modelling classes: tree-based ensemble models, a multi-layer perceptron (MLP), and TabPFN, which was included as a tabular foundation-model baseline. The tree-based ensemble family comprises Random Forest \citep{breiman2001random}, XGBoost \citep{chen2016xgboost}, and LightGBM \citep{ke2017lightgbm}, all operating directly on raw feature values without standardisation. Random Forest aggregates predictions from an ensemble of decision trees, whereas XGBoost and LightGBM build boosted ensembles by sequentially correcting residual errors. These models represent widely adopted off-the-shelf baselines in wildfire susceptibility mapping \citep{marjaniThreeDecadesGeoAI2026} and in downstream applications of AEF embeddings \citep{houAlphaEarthFoundationsAEF2026}.

The neural-network baseline was an MLP implemented with a fixed two-hidden-layer architecture (128 and 64 hidden units, ReLU activations), producing a single output logit trained with binary cross-entropy and optimised with AdamW \citep{loshchilovDecoupledWeightDecay2019}. Finally, TabPFN \citep{hollmannAccuratePredictionsSmall2025} is a transformer-based in-context learner that requires no feature scaling or hyperparameter tuning. We used the TabPFN-3 variant \citep{grinsztajnTabPFN3TechnicalReport2026} as a strong baseline representative of the current state of the art in tabular modelling. The hyperparameter configurations for all tuned models are reported in Table~\ref{tab:hp_tree}.

\paragraph{Spatial neighbourhood models}
To evaluate whether the two feature representations benefited from explicit spatial context, we additionally trained Convolutional Neural Networks (CNNs) using local neighbourhoods centred on each target grid cell. Three patch sizes were considered, $P \in \{9,17,25\}$, corresponding to spatial extents of $9\times9$, $17\times17$, and $25\times25$~km$^2$ at the 1~km grid resolution.

\begin{figure}[htbp]
	\centering
	\includegraphics[width=\textwidth]{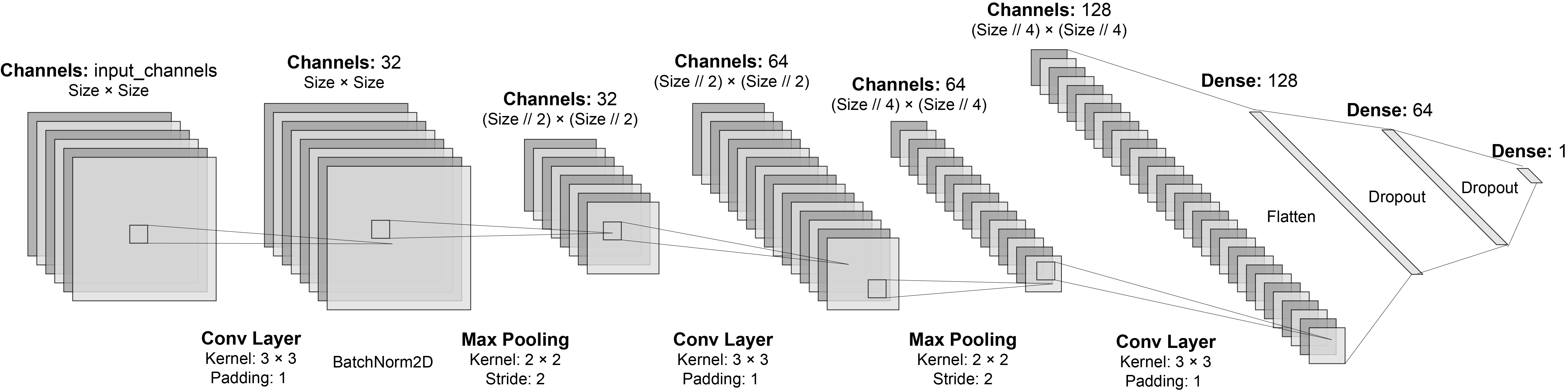}
	\caption{CNN architecture for spatial susceptibility modelling. The convolutional stage extracts spatial features from the $P \times P$ input patch, and the flattened representation is passed to an MLP classifier head with the same structure as the MLP described in Section~\ref{sec:cs_methods}.}
	\label{fig:cnn_arch}
\end{figure}

Each input patch contained $C$ feature channels, where $C=22$ for physical variables and $C=64$ for AEF embeddings. The CNN followed a standard encoder--classifier design, consisting of three convolutional blocks followed by an MLP classifier head (Fig.~\ref{fig:cnn_arch}). Similar CNN-based architectures have been used in previous wildfire susceptibility studies \citep{zhangForestFireSusceptibility2019,bjanesDeepLearningEnsemble2021}. All convolutional layers used $3 \times 3$ kernels with stride~1 and padding~1. Batch normalisation was applied after the first convolutional layer. The first two convolutional layers were followed by $2 \times 2$ max-pooling layers with stride~2, progressively reducing the spatial dimensions of the feature maps. After the final convolutional layer, the resulting feature map was flattened and passed to an MLP classifier head with the same architecture as the MLP described in Section~\ref{sec:cs_methods}. The hyperparameter search space and selected configurations are reported in Table~\ref{tab:hp_cnn_spatial} .

\subsection{Sequence-Informed Susceptibility Modelling}
\label{sec:temporal_methods}

AEF embeddings are natively provided as annual feature vectors, and the physical variables were likewise organised into yearly datasets. In the temporal setting, each grid cell was represented by a sequence of annual feature vectors from $\mathbf{X}^{\text{yr}}$ and $\mathbf{E}^{\text{yr}}$, rather than by the aggregated representations used in the cross-sectional and spatial settings. The response variable, i.e., fire occurrence label, remained cross-sectional, indicating whether the grid cell contained at least one qualified hotspot during 2017--2025. Thus, the temporal models used annual feature sequences to predict the same susceptibility outcome as the other modelling settings.

To incorporate this yearly structure, we modified the MLP and CNN introduced in Section~\ref{sec:cs_methods} by inserting a lightweight yearly weighting mechanism between the 128-unit representation and the remainder of the classifier head, while keeping all other architectural components unchanged. In the modified MLP, each annual feature vector $\mathbf{z}_{i,t}$ was first projected to a 128-dimensional representation through a shared linear layer with ReLU activation:
\begin{equation}
	\mathbf{h}_{i,t}
	=
	\mathrm{ReLU}\!\left(\mathbf{W}\mathbf{z}_{i,t} + \mathbf{b}\right),
	\quad t = 1, \dots, T,
\end{equation}
where $\mathbf{z}_{i,t}$ denotes either $\mathbf{x}_{i,t}$ or $\mathbf{e}_{i,t}$, $\mathbf{W} \in \mathbb{R}^{128 \times C}$ and $\mathbf{b} \in \mathbb{R}^{128}$ are shared across years, and $C$ is the number of input features. The nine yearly representations were then combined through a learnable weighting mechanism. A scalar weight $w_t$ was assigned to each year via softmax over learnable parameters $\boldsymbol{\alpha} \in \mathbb{R}^{T}$:
\begin{equation}
	w_t = \frac{\exp(\alpha_t)}{\sum_{k=1}^{T} \exp(\alpha_k)},
	\qquad
	\bar{\mathbf{h}}_i = \sum_{t=1}^{T} w_t \,\mathbf{h}_{i,t}.
\end{equation}
The pooled representation $\bar{\mathbf{h}}_i \in \mathbb{R}^{128}$ was then passed to the remainder of the classifier head ($128 \to 64 \to 1$). The learned weights $\{w_t\}$ were retained for interpretation, indicating the relative contribution of each year to the final susceptibility prediction.

The modified CNN applied the same weighting mechanism at the same position in the architecture. Each year's $P \times P$ patch was processed by the shared convolutional backbone, flattened, and projected to a 128-dimensional spatial feature vector. The nine yearly spatial vectors were then aggregated using the learned year weights before being passed to the remaining classifier head.

Both modified architectures were evaluated under the physical and AEF embedding settings. For the CNN-based models, the same three patch sizes $P \in \{9,17,25\}$ were used. The hyperparameter search spaces and selected configurations are reported in Table~\ref{tab:hp_temporal} .

\subsection{Reconstruction of Physical Variables}\label{sec:reconstruction}

Previous modelling evaluated whether AEF embeddings contain wildfire-relevant information, but did not identify which environmental signals were encoded. To address this, we conducted a reconstruction analysis that probed the extent to which individual physical variables could be recovered from the embeddings. For each calendar year $t \in \{2017, \dots, 2025\}$ and each continuous variable $v \in \mathcal{V}_{\mathrm{cont}}$\footnote{For the reconstruction of land-cover variables, readers are referred to \citet{benavides-martinezWhatEarthAlphaEarth2026}.}, a separate year-specific regression model was trained to predict the observed value of variable $v$ from the 64-dimensional AEF embedding of the same year:
\begin{equation}
	\hat{x}_{i,t}^{(v)} = f_{v,t}\bigl(\mathbf{e}_{i,t}\bigr),
\end{equation}
where $f_{v,t}$ is a single-hidden-layer MLP with 32 hidden units trained to minimise mean squared error using the Adam optimiser, with a learning rate of $10^{-2}$, dropout rate of 0.1, and 800 training epochs, determined via a preliminary search. The same training and test split defined in Section~\ref{sec:overview} was used. Reconstruction quality is reported as the distribution of test-set $R^{2}$ values across the nine years for each variable.

\subsection{Evaluation Framework}\label{sec:evaluation}
\subsubsection{Within-Victoria Performance Evaluation}  \label{sec:victoriaevaluation}

Model performance was evaluated on the test set within Victoria. Five binary classification metrics \citep{JamesWitten-81} are reported at a decision threshold of 0.5: Accuracy, Recall, Specificity, F1-score, and ROC-AUC. Accuracy measures overall correctness, while recall and specificity evaluate the ability to identify positive and negative samples, respectively. F1-score balances precision and recall, which is particularly relevant in wildfire susceptibility modelling where both false negatives and false positives carry risks. Let $\TP$, $\FP$, $\TN$, and $\FN$ denote true positives, false positives, true negatives, and false negatives, respectively. The threshold-dependent metrics are defined as:
\begin{align}
	\mathit{Accuracy} &=
	\frac{\TP + \TN}{\TP + \TN + \FP + \FN}, \\
	\mathit{Recall} &=
	\frac{\TP}{\TP + \FN}, \\
	\mathit{Specificity} &=
	\frac{\TN}{\TN + \FP}, \\
	\mathit{F1\text{-}score} &=
	\frac{2 \times \mathit{Precision} \times \mathit{Recall}}
	{\mathit{Precision} + \mathit{Recall}},
\end{align}
where precision is used only to compute F1-score and is defined as
\begin{equation}
	\mathit{Precision} =
	\frac{\TP}{\TP + \FP}.
\end{equation}

ROC-AUC measures the model's overall discriminative capacity across thresholds. Its computation follows \citet{jinhuangUsingAUCAccuracy2005}.

\subsubsection{Transferability Assessment}\label{sec:transfer_methods}
To assess whether the learned relationships generalised beyond the training region, the model trained under each of the 10 independent Victoria train/test splits was applied, without modification, to eight transfer sites across Australia (Section~\ref{sec:study_area_transfer}). At each transfer site, a balanced dataset was constructed at a spatial resolution of 1~km following the same procedure described in Sections~\ref{subsec:hand}--\ref{subsec:fire}, and the sample counts and study area extents for each transfer site are summarised in Table~\ref{tab:transfer_sites}.

The relative transfer performance change is quantified as
\begin{equation}
	\Delta_m^{\,s} = \frac{M_m^{\,s} - M_m^{\,\text{VIC}}}{M_m^{\,\text{VIC}}},
    \label{eq:transfer_delta}
\end{equation}
where $M_m^{\,s}$ denotes the mean value of metric $m$ evaluated at transfer site $s$ across the models trained under the 10 independent Victoria train/test splits, and $M_m^{\,\text{VIC}}$ denotes the mean value of the corresponding metric on the Victoria test sets across the same 10 runs. Negative values of $\Delta_m^{\,s}$ denote performance degradation at the transfer site relative to the source region.

For conciseness, transfer results are reported using aggregated metrics. Models were grouped into four families: tree-based ensembles, MLP, TabPFN, and CNN. For each transfer site and metric, $\Delta_m^{\,s}$ was first averaged within each model family and then averaged across the four families, yielding the final cross-family mean performance change.

\section{Results} \label{sec:results}

\subsection{Susceptibility Modelling}
\label{sec:results_performance}

\begin{figure}[htbp]
	\centering
	\includegraphics[width=0.8\linewidth]{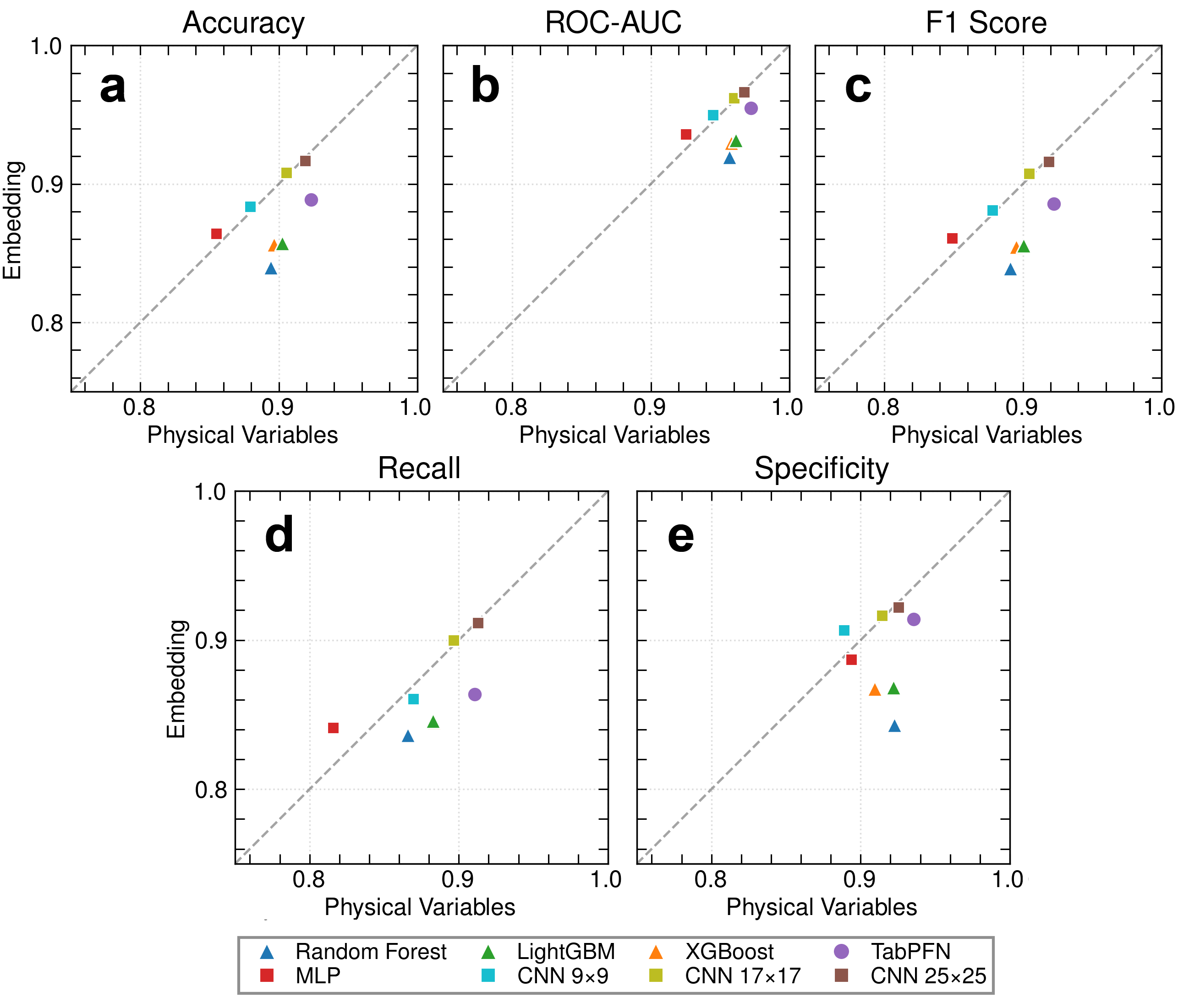}
	\caption{Comparison of time-aggregated model performance between physical variables and AEF embeddings in Victoria test set. Points above the dashed $y = x$ line correspond to higher performance for AEF embeddings, while points below the line correspond to higher performance for physical variables.}
	\label{fig:benchmark_embedding}
\end{figure}

Both AEF embeddings and hand-engineered physical variables achieve strong within-Victoria classification performance on the test set, with ROC-AUC values generally above 0.92 across the time-aggregated settings (Fig.~\ref{fig:benchmark_embedding}), with standard deviations in ROC-AUC below $0.005$ across all models, while accuracy, F1 score, recall and specificity are all above 0.80 and reach approximately
0.90 at their upper end (Fig.~\ref{fig:benchmark_embedding}, Table~\ref{tab:metric_mean_sd}). The relative performance of the two feature representations varies by model family. Tree-based ensemble models favour physical variables over AEF embeddings, and their feature-importance profiles assign higher ranks to structurally stable predictors (e.g., solar radiation, wind speed, precipitation and grass land cover) than to fire-season persistence and extreme-condition variables, denoted by F.S. in Table~\ref{tab:variables}  (Fig.~\ref{fig:physical_feature_importance}). In the embedding-based tree models, the most important predictors are concentrated in a stable subset of embedding dimensions (Table~\ref{tab:embedding_top_dims}). By contrast, the MLP, CNN, and TabPFN results are positioned closer to the $y=x$ line, indicating more comparable performance between the two feature representations under flexible susceptibility models.

CNN-based models that use local neighbourhood patches inprove performance over the cell-wise MLP baseline for both feature representations, although the additional gain decreases as the neighbourhood size increases. Moving from a $9 \times 9$ to a $17 \times 17$ input patch produces clearer improvements than moving from $17 \times 17$ to $25 \times 25$. With AEF embeddings, CNN with $17 \times 17$ and $25 \times 25$ patch inputs outperforms the tabular foundation-model baseline TabPFN across all five reported metrics. With physical variables, CNN models do not exceed the strongest tabular baselines. The marginal gain from explicit spatial modelling is larger for AEF embeddings than for hand-engineered physical variables.

\begin{figure}[htbp]
	\centering
	\includegraphics[width=0.7\linewidth]{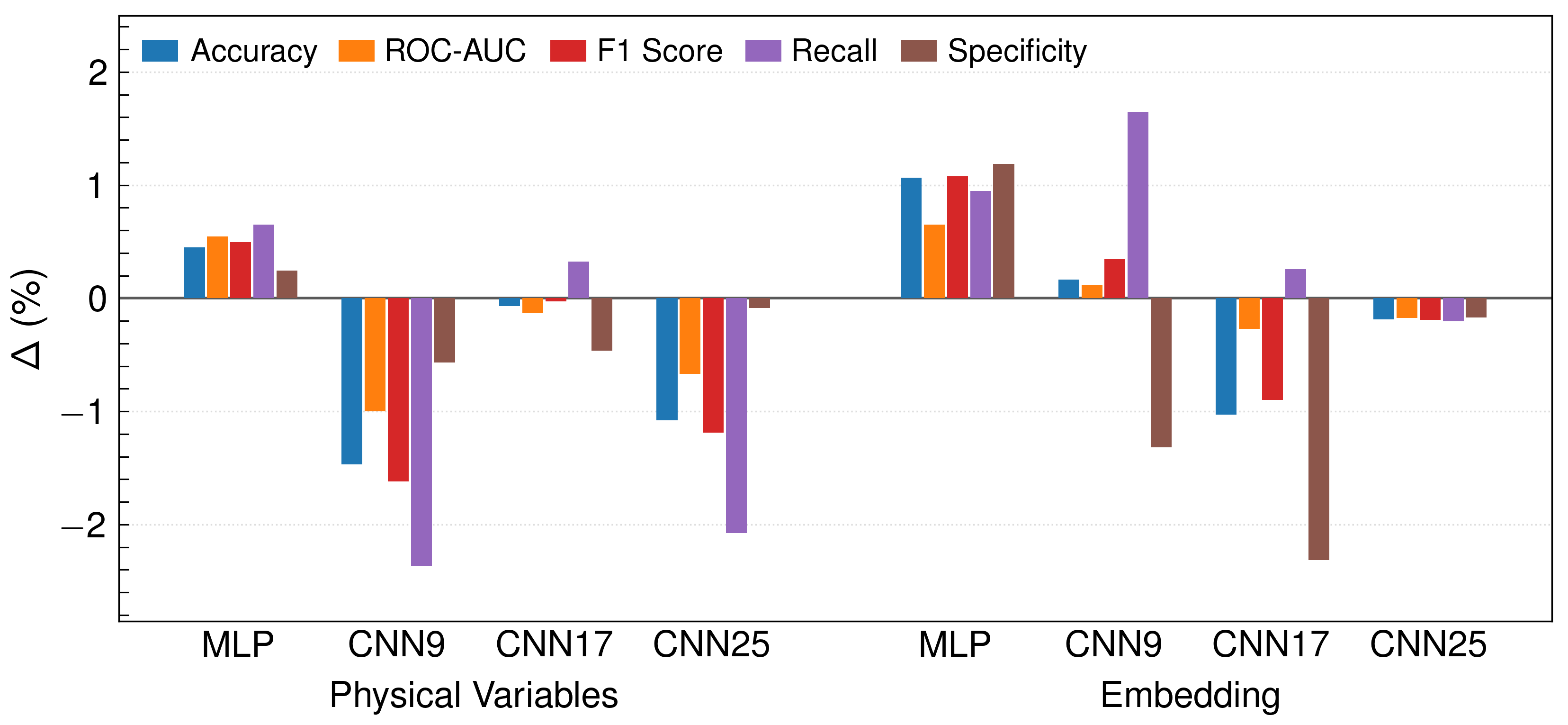}
	\caption{Performance change of temporal models relative to their corresponding non-temporal counterparts. Values are expressed as percentage-point changes.}
	\label{fig:temporal_gain}
\end{figure}

Sequence-informed modelling produces only modest performance changes relative to the corresponding non-temporal counterparts (Fig.~\ref{fig:temporal_gain}). The largest positive changes are observed for the MLP models, but most improvements are below 1.5\%. Most CNN-based temporal models show small negative or near-zero changes, particularly under the physical-variable representation. Despite the limited performance gains, the learned temporal weights are not uniformly distributed across years. Several models assign unusually high weights to 2020 relative to other years (Fig.~\ref{fig:temporal_weights}).

\subsection{Reconstruction Analysis}\label{sec:results_reconstruction}

\begin{figure}[htbp]
	\centering
	\includegraphics[width=0.5\linewidth]{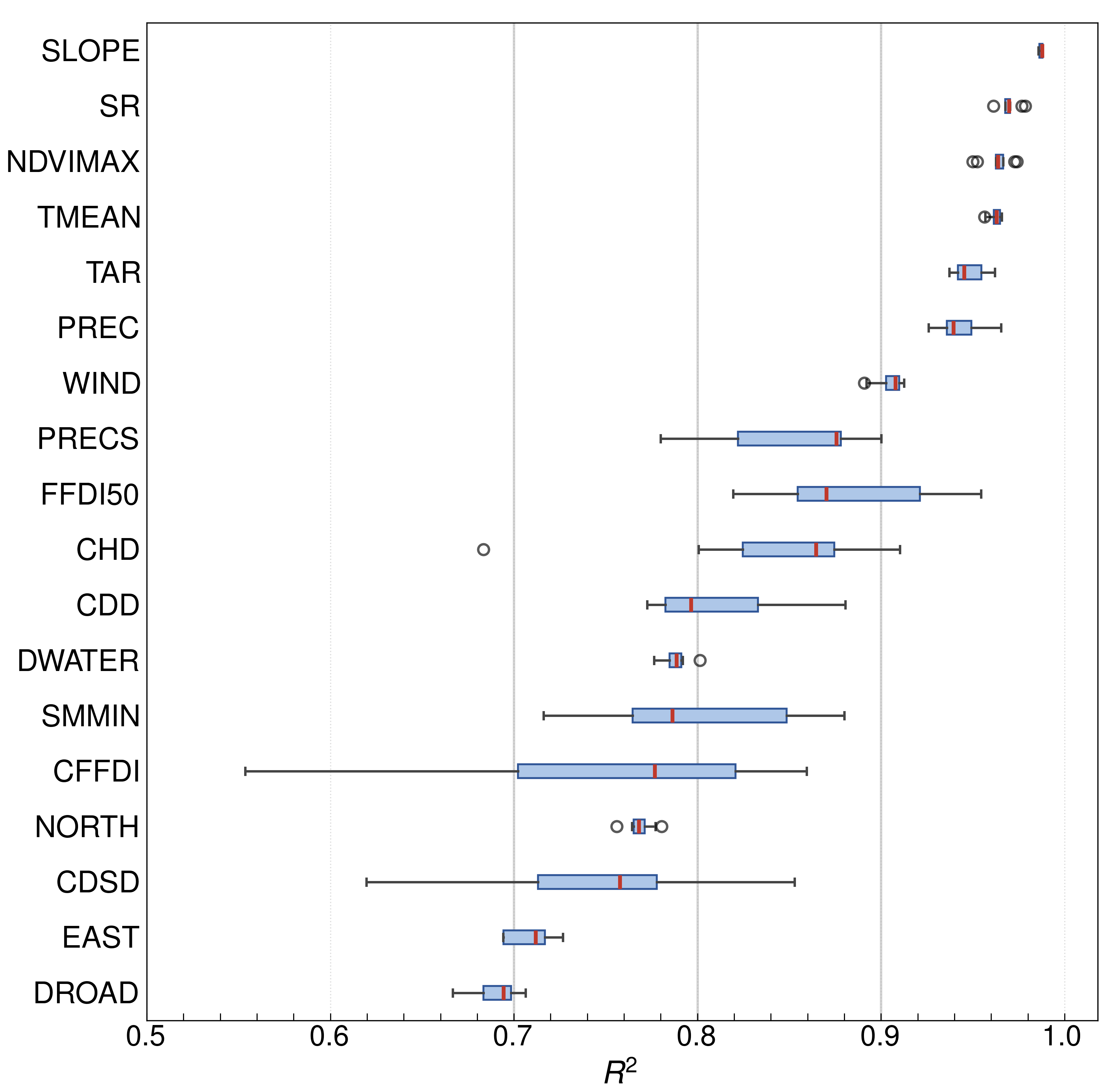}
	\caption{Reconstruction performance of selected physical variables from annual AEF embeddings. Boxplots summarise yearly $R^2$ values across 2017--2025.}
	\label{fig:reconstruction_r2}
\end{figure}

The reconstruction analysis indicates which physical variables can be recovered from AEF embeddings (Fig.~\ref{fig:reconstruction_r2}). Overall, many variables are reconstructed with median $R^2$ values above 0.8, with several predictors achieving $R^2$ values close to or above 0.9 across years. Slope, solar radiation, maximum NDVI, mean temperature, temperature annual range, total precipitation, and mean wind speed show particularly strong and stable reconstruction performance. These well-reconstructed variables largely overlap with the physical predictors assigned high feature importance in the tree-based models (Fig.~\ref{fig:physical_feature_importance}) and correspond to commonly used predictors in wildfire susceptibility mapping \citep{xuDeepLearningWildfire2025}.

By comparison, derived directional and proximity variables, including northness, eastness, and distance to road, had consistently lower \(R^2\) values across years. Reconstruction was also less temporally stable for persistence and extreme-condition summaries computed over the fire season, including days with FFDI above 50, consecutive hot days, consecutive dry days, and consecutive dry-soil days, whose \(R^2\) values varied substantially among years.

\subsection{Wildfire Susceptibility Mapping in Victoria}
\label{sec:results_mapping}

\begin{figure}[htbp]
	\centering
	\includegraphics[width=0.9\linewidth]{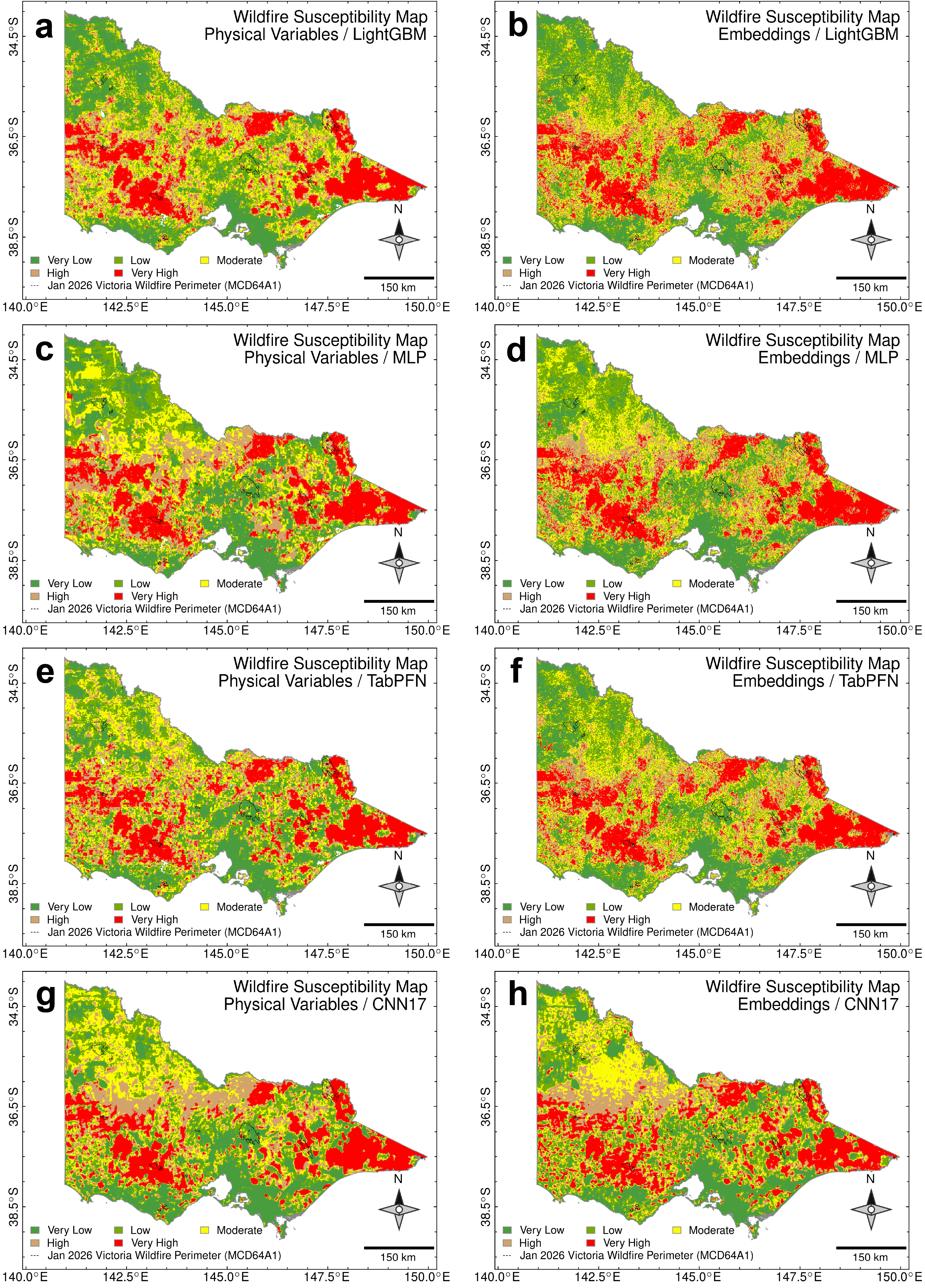}
	\caption{Wildfire susceptibility maps for representative physical-variable and embedding-based models in Victoria.}
	\label{fig:composite_map}
\end{figure}

After model training, the fitted models were applied to all 1~km grid cells across Victoria. For each cell, the model output for the positive fire-occurrence class was used as a continuous susceptibility score\footnote{Although bounded between 0 and 1, these scores should not be interpreted as absolute ignition probabilities. Because training used a 1:1 positive-to-negative ratio, the scores reflect relative spatial propensity for ignition rather than true occurrence rates. Converting them to absolute probabilities requires calibration against the observed ignition base rate \citep{depickerWildfireIgnitionProbability2020}.}. For map comparison, the scores were converted into five percentile-based \footnote{Jenks natural breaks \citep{jenksDataModelConcept1967} were not used because the method optimises class boundaries according to the score distribution of each model independently, which can produce inconsistent class area proportions across models. Percentile-based thresholds enforce identical area proportions for each class across all models, enabling direct comparison of spatial agreement in high-susceptibility classifications.} susceptibility classes, with the top 20\% classified as very high susceptibility, the 20--40\% percentile range as high susceptibility, and the remaining ranges classified accordingly. To provide an out-of-period qualitative reference, the susceptibility maps were overlaid with the burned perimeter derived from the NASA MCD64A1 product \citep{giglioMODISTerraAquaBurned2021} for the January 2026 Victorian bushfires \citep{vicgov2026bushfires}. Fig.~\ref{fig:composite_map} presents the representative maps, using LightGBM for the tree-based models and the 17 $\times$ 17 CNN for the spatial models. The full set of tree-based and CNN maps is provided in Figs.~\ref{fig:composite_trees} and~\ref{fig:composite_cnn}, respectively.

The susceptibility maps show broadly consistent spatial structure across model families and feature representations (Fig.~\ref{fig:composite_map}). High and very high susceptibility classes are repeatedly concentrated across inland Victoria, including parts of southwestern and central Victoria as well as the forested and mountainous landscapes of eastern Victoria associated with the Great Dividing Range, the Victorian Alps, and Gippsland. Lower susceptibility is assigned more consistently to the Port Phillip Bay surrounds, parts of the southern coastal plain, and highly urbanised areas. The January 2026 burned perimeter is overlaid in Fig.~\ref{fig:composite_map} as an out-of-period qualitative reference.

One clear divergence in model behaviour occurs in northwestern Victoria, which has a distinctly different climate from the southern regions (Fig.~\ref{fig:study_area}a) and where fire occurrences are spatially isolated. Most models capture this pattern of isolated high-susceptibility patches, with the exception of the MLP (Fig.~\ref{fig:composite_map}c, d). Differences are also evident in spatial texture. Tree-based and TabPFN maps are relatively fine-grained, with high and very high susceptibility pixels interspersed among moderate and lower-susceptibility areas (Fig.~\ref{fig:composite_map}a, b, e, f). By contrast, the CNN maps produce a more spatially aggregated pattern, where high and moderate susceptibility classes form contiguous zones (Figs.~\ref{fig:composite_map}g, h and~\ref{fig:composite_cnn}).

In addition to binary classification, the predicted susceptibility scores from the cell-level and spatial neighbourhood models also reflect fire-recurrence patterns, showing a monotonic relationship with the number of months in which MODIS fire occurrences were recorded during 2017--2025 (Figs.~\ref{fig:scorevsfirecountphysical} and~\ref{fig:scorevsfirecountpanelembedding}). This pattern is observed for both physical-variable and embedding-based models, with cells experiencing more fire-occurrence months receiving higher mean susceptibility scores.

\subsection{Transferability across Australia}
\label{sec:results_transfer}

\begin{figure}[htbp]
	\centering
	\includegraphics[width=0.8\linewidth]{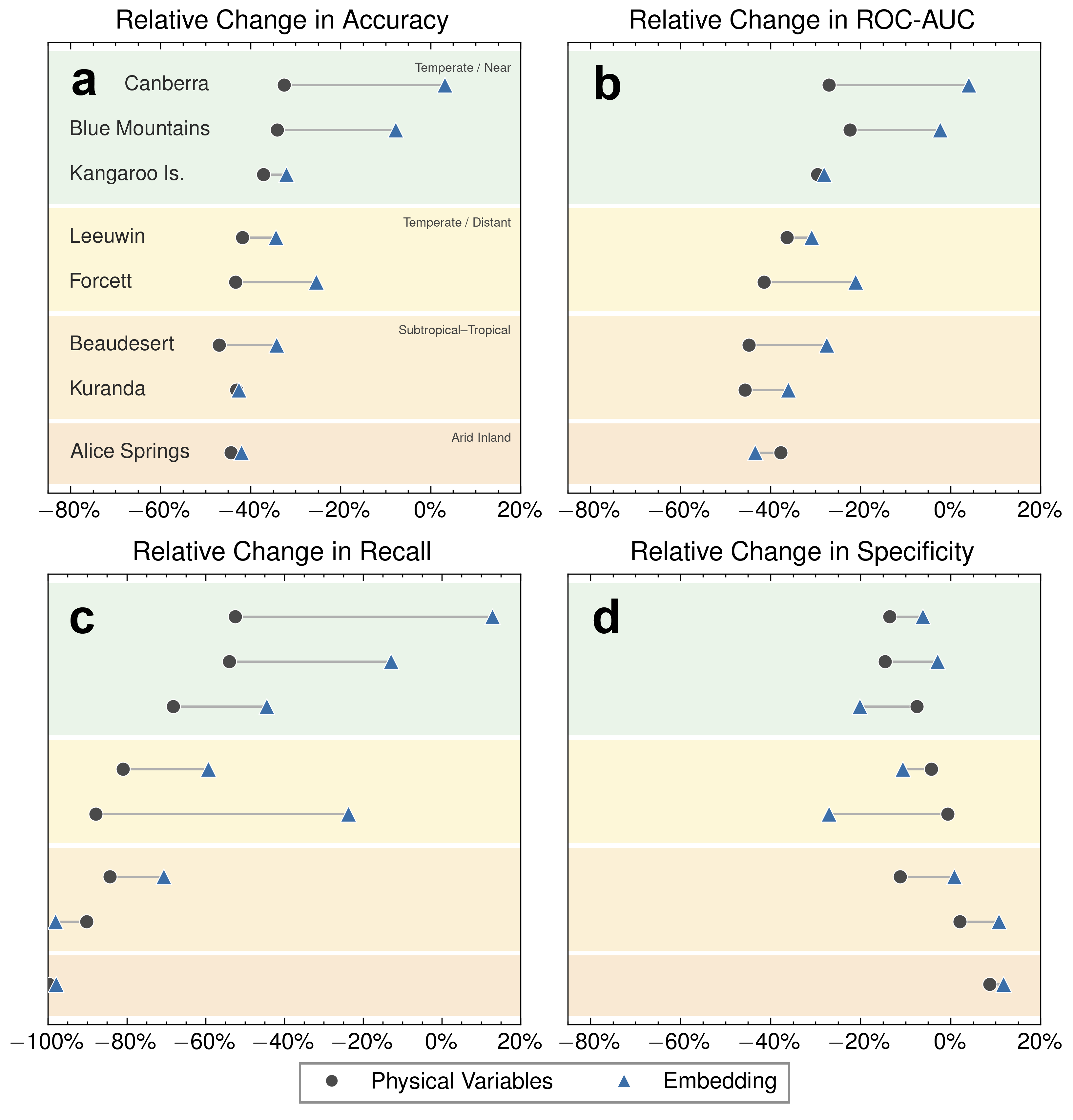}
	\caption{Mean transfer performance change relative to Victoria. Negative values denote performance degradation at the transfer site. Values are averaged across model families within each feature representation.}
	\label{fig:transfer_delta}
\end{figure}

Transferability declines as target regions become more climatically distant from Victoria and less similar in fire regime, but the form of this decline differs markedly between physical-variable and embedding-based models (Figs.~\ref{fig:transfer_delta} and~\ref{fig:transfer_absolute}). Physical-variable models exhibit abrupt degradation in accuracy and ROC-AUC once applied outside Victoria: even at climatically related temperate sites such as Canberra and Western Sydney--Blue Mountains, their ROC-AUC drops by an average of approximately 25\%. Embedding-based models, by contrast, show a much more gradual decline. At Canberra, ROC-AUC improves by around 4\% on average, and recall increases substantially (by around 13\% on average). At Western Sydney--Blue Mountains, their mean ROC-AUC decline is only around 2\%. This advantage remains highly stable across the 10 independent runs (Table~\ref{tab:transfer_delta_sd}). Performance losses increase at more distant temperate sites such as Kangaroo Island, Leeuwin--Boranup, and Forcett--Dunalley. Larger degradation appears in the subtropical--tropical sites in Queensland and is more pronounced in the arid inland setting around Alice Springs. To summarise, AEF embeddings show stronger and more gradual cross-regional transfer than physical variables, particularly in climatically similar regions. However, their performance still declines in more distant and climatically distinct environments, indicating that transferability remains conditional rather than universal.

As transfer distance and climatic dissimilarity increase, recall degrades more rapidly than specificity (Fig.~\ref{fig:transfer_delta}c, d; Fig.~\ref{fig:transfer_absolute}). Correspondingly, the relative advantage of embedding-based models at near temperate sites is mainly reflected in their better retention of the ability to identify positive (fire-occurrence) pixels.

\section{Discussion}\label{sec:discussion}
\subsection{Wildfire-Relevant Information in AEF Embeddings}

AEF embeddings are increasingly used as analysis-ready geospatial representations to substitute for physical variables when such variables are unavailable, difficult to harmonise, or costly to engineer. As a commonly used risk-assessment mapping task traditionally requiring a substantial set of multi-source physical variables, wildfire susceptibility mapping may also benefit from such representations. However, their applicability to this task, and the extent to which they contain wildfire-relevant information, remain insufficiently evaluated.

Our experiments in Victoria, Australia demonstrate AEF embeddings contain substantial wildfire-relevant information. In the cell-wise setting, where each grid cell is represented by its own feature vector, embedding-based models achieve ROC-AUC values close to those of carefully engineered physical-variable models, indicating much of the discriminative signal required for susceptibility mapping is already present in the embedding space. When local neighbourhood information is incorporated through CNN-based patch models, the two representations become fully comparable. The architecture-specific explanation for this improvement is discussed separately in Section \ref{subsec:archi}. 

The resulting embedding-based maps successfully identify the high-susceptibility regions of eastern and southwestern Victoria documented in prior literature \citep{hosseiniGeneExpressionProgramming2021}, and most models also capture isolated high-susceptibility patches in the climatically distinct northwestern part of the state. Their visual overlap with the January 2026 burned perimeter, derived from MODIS burned-area data, provides an out-of-period plausibility check. This overlap should not be interpreted as formal temporal validation, but it suggests that the learned susceptibility patterns are not merely artefacts of the 2017–2025 fire-occurrence inventory. 

Furthermore, our reconstruction analysis clarifies why embedding-based susceptibility models perform competitively. Many variables assigned high feature-importance scores by the tree-based susceptibility models, including vegetation condition, solar radiation, temperature, precipitation, and wind, can be recovered from the 64-dimensional AEF representation with high accuracy. This extends the findings of \citet{rahmanPhysicallyInterpretableAlphaEarth2026} from US data to an Australian, fire-specific setting, and confirms AEF embeddings preserve environmental structure relevant to wildfire occurrence despite not being specially trained for this task. The same logic may extend to other natural hazards whose susceptibility is shaped by persistent land-surface conditions, such as flood and landslide susceptibility \citep{islamSystematicReviewUrban2025,luuIntegratingSusceptibilityMaps2024,reichenbachReviewStatisticallyBased2018}. Whether this potential extends beyond wildfire susceptibility remains an important direction for future research.

The strong performance of AEF embeddings also suggests a practical use as a meaningful performance baseline, even for researchers who ultimately adopt physical-variable models for interpretability. Wildfire susceptibility mapping is not a fully standardised machine learning task, and no established benchmark currently accounts for the diversity of study regions, spatial resolutions, and data processing choices found across the literature. Because AEF embeddings are analysis-ready, researchers can construct a study-specific baseline and evaluate it quickly, without first committing to a full physical-variable pipeline. When a physical-variable model fails to exceed this baseline, the gap signals a need to revisit variable selection, refine data processing, or incorporate additional predictors. Further guidance on which predictors to prioritise may come from ongoing research into AEF's training data and the information content of its representations \citep{benavides-martinezWhatEarthAlphaEarth2026,rahmanPhysicallyInterpretableAlphaEarth2026}.

\subsection{Architecture Matters for Extracting AEF Susceptibility Signal } \label{subsec:archi}

Beyond evaluating whether AEF embeddings can be a useful alternative for hand-engineered physical variables, this study also contributes an architecture-level comparison of how wildfire-relevant information is extracted from the embedding field. The results show that cell-wise tabular models already access substantial signal from AEF embeddings, but CNN-based patch models extract additional predictive information by using local neighbourhood context. This distinction is important because evaluations based only on off-the-shelf tabular learners may understate the practical value of gridded geospatial embeddings. More generally, our findings suggest that foundation-model embeddings should be evaluated not only as tabular covariates, but also as spatially structured fields that can be used by computer-vision-style architectures. This implication should be interpreted cautiously, however, because the result is task- and scale-dependent and should be tested across additional regions and hazard-mapping applications. 

This pattern is consistent with the structure of the embedding field itself. In the Space-Time-Precision (STP) encoder of AlphaEarth Foundations, the space operator uses a Vision-Transformer-like spatial self-attention mechanism to capture broader spatial dependencies, while the precision operator uses $3 \times 3$ convolutional operations to maintain local spatial detail \citep{brownAlphaEarthFoundationsEmbedding2025}. These design features suggest that AEF embeddings are not simply independent cell-wise vectors, but part of a spatially structured embedding field containing both neighbourhood-level and broader contextual information. The stronger performance of CNN-based patch models therefore suggests that some wildfire-relevant signal is expressed through local spatial structure in the embedding field, which cell-wise tabular models cannot directly use. This helps explain why larger patches, particularly the $17 \times 17$ and $25 \times 25$ patches, were especially effective for AEF embeddings in our task. 

The use of neighbourhood context also explains why the CNN maps form smoother and more spatially contiguous susceptibility zones than the cell-level maps. Such spatial coherence may be useful for susceptibility zoning and regional planning, where contiguous zones can provide more actionable management units than isolated high-susceptibility pixels. However, this smoothing may also obscure small localised hotspots, so its practical value depends on the intended use of the susceptibility map.

By contrast, annual-sequence modelling added limited predictive benefit. This likely reflects the structure of the prediction task as much as the data resolution. Although the models receive annual feature vectors, the response variable remains a single 2017--2025 fire-occurrence label, so the model is still estimating overall susceptibility rather than year-specific fire occurrence. If susceptibility is dominated by persistent land-surface, vegetation, topographic, and climatic gradients, the time-aggregated representation may already contain most of the relevant predictive signal. Aggregating 10 m AEF embeddings to the 1 km modelling grid may further reduce fine-scale annual variation, particularly where year-to-year changes occur at sub-grid scales. This contrasts with change-detection applications, which often operate at the native 10~m resolution to use annual embedding layers for characterising changes over time \citep{chaiyanaScalablePotentialAlpha2026}. The limited gain from annual-sequence modelling therefore does not imply annual AEF embedding layers lack temporal information. Rather, it suggests that such information is difficult to exploit when the embeddings are aggregated to 1~km and used to predict a cross-sectional outcome.

The elevated weight assigned to 2020 should be interpreted cautiously as a predictive association, rather than evidence that 2020 conditions were causally more important for fire susceptibility. One likely explanation is the concentration of fire-occurrence labels associated with the 2019--2020 Black Summer fires  \citep{abramConnectionsClimateChange2021} with approximately 28\% of fire occurrence grid cells in our inventory originating from 2020. In addition, the environmental conditions of 2019--2020 were anomalous across multiple climate indicators \citep{devanandAustraliasTinderboxDrought2024,vanoldenborghAttributionAustralianBushfire2021}, and these signals may be partly encoded in AEF embeddings through their use of ERA5-derived information. Therefore, the high 2020 weight suggests that this annual layer was especially informative for distinguishing burned from unburned cells in this study period. A sensitivity analysis excluding 2020, or separating pre-fire from post-fire information, would be needed to test this interpretation further, but this was beyond the scope of the present study.

\subsection{Spatial Transferability and Scalable Susceptibility Mapping}
In wildfire susceptibility mapping, model generalisation is typically assessed using randomly held-out test samples from the same study region, whereas spatial transferability across distinct regions has received comparatively little attention. At Canberra, the transfer site geographically closest to Victoria (Fig.~\ref{fig:study_area}a), models trained on physical variables in Victoria exhibit a substantial decline in recall and ROC-AUC. This result aligns with \citet{bekarCrossregionalModellingFire2020}, who found that most cross-regional transfers of regional fire-occurrence models in the Alps and the Mediterranean Basin were unsuccessful, with southern Switzerland to Carinthia, Austria, as the main exception. In our study, AEF model performance also declined outside Victoria, but less sharply than physical-variable model performance in nearby regions within similar climate zones. One possible explanation is that AEF was trained on a global, multi-billion-observation corpus spanning diverse climates and land-surface conditions and produces a unified embedding field over the global land surface, within which environmentally similar regions may be represented in a more comparable form. This advantage may diminish as source--target distance and environmental dissimilarity increase, because the relationships between embedding features and downstream labels can vary across regions \citep{yangEvaluatingPerformanceAlphaEarth2026}. Consistent with this possibility, agricultural studies involving transfers from the United States to Argentina and from China to the United States have reported limited AEF transferability \citep{maHarvestingAlphaEarthBenchmarking2026,yangEvaluatingPerformanceAlphaEarth2026}. Future research could determine how this transfer advantage varies with source--target distance, environmental similarity, and downstream task, and identify the conditions under which it disappears.

The conditional transfer pattern observed here suggests practical implications for large-area susceptibility mapping across diverse climate zones, which is relevant to public agencies managing wildfire exposure across broad jurisdictions and to insurers and reinsurers whose portfolios span multiple regions or entire countries. Where physical-variable models transfer poorly, reliable large-area mapping may require labelled samples across much of the target domain to adequately represent local environmental relationships, resulting in large and spatially extensive training datasets, particularly in countries with diverse climate zones and fire regimes. Embedding-based modelling may offer a more scalable alternative: representative areas could be sampled within major climate or fire-regime zones, and models trained in these areas could then be transferred to environmentally similar regions with more controlled degradation.

\subsection{Annual Geospatial Embeddings as Susceptibility Predictors }

AEF annual layers are better understood as reusable representations of yearly land-surface and environmental conditions than as substitutes for long multi-decadal climate records. This design is well suited to wildfire susceptibility mapping, where the objective is to estimate relative spatial propensity for fire occurrence from vegetation, land-surface, topographic, and climatic context rather than to forecast short-term ignition events. By summarising conditions over a full year, AEF embeddings can capture persistent and seasonal landscape characteristics that are directly relevant to susceptibility assessment. 
At the same time, the scientific community would benefit from continued annual extension of the AEF product, and where feasible, longer retrospective archives. Related remote-sensing foundation-model products such as TESSERA \citep{fengTESSERATemporalEmbeddings2025} and Prithvi-EO-2.0 \citep{szwarcmanPrithviEO20VersatileMultiTemporal2025} similarly provide archives covering roughly the last 10--15 years. A longer archive would support stronger retrospective evaluation \citep{sunFirstMapPlanted2026}, improve separation between persistent susceptibility patterns and short-period climatic anomalies, and enable more robust fire-climate analyses.  

The reconstruction of variables defined by fire-season persistence or short-term extremes, including days with FFDI above 50, consecutive hot days, consecutive dry days, and consecutive dry-soil days, are more variable across years. Therefore, for near-real-time ignition prediction or next-day fire-risk forecasting,  AEF embeddings are best used as environmental context alongside dynamic meteorological forcings, rather than as standalone predictors \citep{zlydenkoAIExpandsHighquality2026}.

\section{Conclusion}\label{sec:conclusion}

This study systematically evaluates AlphaEarth Foundations (AEF) embeddings for wildfire susceptibility mapping. Using Victoria, Australia, as the source region, we show that AEF embeddings contain substantial wildfire-relevant information and can support susceptibility models with performance comparable to physical-variable models, particularly when downstream models use neighbouring information through patch-based input. Transfer experiments further show embedding-based Victorian models retain stronger performance in nearby temperate regions, especially Canberra and Western Sydney--Blue Mountains, than models based on physical variables. Overall, our findings highlight the potential of AEF embeddings for wildfire susceptibility mapping tasks dominated by structurally stable environmental information. Future research could extend this evaluation to larger spatial domains and to other hazards, such as floods and landslides, to assess whether geospatial embeddings can serve as reusable data infrastructure for scalable multi-hazard susceptibility assessment.

\section*{Acknowledgements}
This work was supported by Australian Research Council’s Discovery Projects funding scheme (DP250104816). Y.Z., S.H., and F.H. acknowledge the support of the Australian Research Council Centre of Excellence for the Weather of the 21st Century (CE230100012). The work was undertaken using resources from the National Computational Infrastructure (NCI), which is supported by the Australian Government. The authors acknowledge Patrick Laub, Bernard Wong, Len Patrick Garces, Gregory Taylor, Jinxia Zhu and Gerry (Zherui) Li for their helpful suggestions.

\section*{Data availability}
In our research, the data used are publicly available, and the corresponding references to the datasets are provided in the manuscript.

\bibliographystyle{plainnat}
\bibliography{references}
\clearpage
\appendix
\clearpage
% Save the main-article page total before supplementary numbering restarts.
\xdef\MainLastPage{\number\numexpr\value{page}-1\relax}

% Supplementary Material included by Main Text 20260615.tex.
% It shares the main document's labels, but its citations and reference list
% are isolated by bibunits.
\begingroup

% Independent supplementary page, section, and float numbering.  The
% supplementary body uses the same geometry and typography as the main text.
\setcounter{page}{1}
\renewcommand{\thepage}{\arabic{page}}
\setcounter{section}{0}
\renewcommand{\thesection}{S\arabic{section}}
\renewcommand{\theHsection}{supp.\arabic{section}}
\setcounter{figure}{0}
\renewcommand{\thefigure}{S\arabic{figure}}
\renewcommand{\theHfigure}{supp.\arabic{figure}}
\setcounter{table}{0}
\renewcommand{\thetable}{S\arabic{table}}
\renewcommand{\theHtable}{supp.\arabic{table}}
\setcounter{equation}{0}
\renewcommand{\theequation}{S\arabic{equation}}
\renewcommand{\theHequation}{supp.\arabic{equation}}
\pagestyle{supplementary}
\thispagestyle{supplementaryfirst}

% Compact arXiv-style supplementary title block.
\begin{center}
  \rule{\linewidth}{1.2pt}\par
  \vspace{0.8em}
  {\LARGE\bfseries Supplementary Material\par}
  \vspace{0.5em}
  {\large\bfseries Evaluating AlphaEarth Foundations Embeddings for Wildfire Susceptibility Mapping\par}
  \vspace{0.8em}
  \rule{\linewidth}{1.2pt}\par
\end{center}
\vspace{1.1em}

% This unit has its own bibliography and does not add citations to the
% reference list of the main article.
\begin{bibunit}[plainnat]

\section{Variable Quantification and Visualization}\label{sec:app_variables}

This Supplementary Material documents the computational definitions of all 35 candidate predictor variables listed in main text Table~\ref{tab:variables}, as well as the visualization of the cross-sectional dataset.

% -------------------------------------------------------
\subsection{Topographic Variables (Fig.~\ref{fig:topographic})}\label{subsec:topo}
\vspace{0.1cm}
\noindent All topographic variables are static (time-invariant).

\vspace{0.1cm}
\noindent \textbf{1 -- Elevation} (ELEV):
\begin{itemize}
	\item \textit{Unit:} m
	\item \textit{Calculation:} Mean elevation within each 1~km cell.
\end{itemize}

\noindent \textbf{2 -- Slope} (SLOPE):
\begin{itemize}
	\item \textit{Unit:} $^\circ$
	\item \textit{Calculation:} Computed using central-difference gradients via \texttt{numpy.gradient()}:
	\begin{equation}
		\text{SLOPE} = \arctan\!\sqrt{g_x^2 + g_y^2},\qquad (g_x, g_y) = \nabla z
	\end{equation}
	Averaged to 1~km grid.
\end{itemize}

\noindent \textbf{3--4 -- Northness} (NORTH) and \textbf{Eastness} (EAST):
\begin{itemize}
	\item \textit{Unit:} NA
	\item \textit{Calculation:} Slope aspect (the compass direction a slope face points towards) influences wildfire susceptibility through its control on solar radiation receipt and vegetation structure. In the Southern Hemisphere, north-facing slopes receive more direct insolation, leading to drier and more fire-prone conditions \citep{lindenmayerEmpiricalAnalysesFactors2021}. Aspect is represented by northness and eastness to avoid treating 0$^\circ$ and 360$^\circ$ as numerically distant values:
	\begin{equation}
		\text{NORTH} = \cos\!\left(\alpha \cdot \frac{\pi}{180}\right),\qquad
		\text{EAST}  = \sin\!\left(\alpha \cdot \frac{\pi}{180}\right)
	\end{equation}
	Both components range $[-1, +1]$. NORTH $= +1$ indicates a due-north-facing slope, NORTH $= -1$ a due-south-facing slope, and NORTH $= 0$ for an east- or west-facing slope. EAST is interpreted analogously. Both NORTH and EAST are computed at the native DEM resolution and subsequently averaged to 1~km grid.
\end{itemize}

\begin{figure}[htbp]
	\centering
	\includegraphics[width=\linewidth]{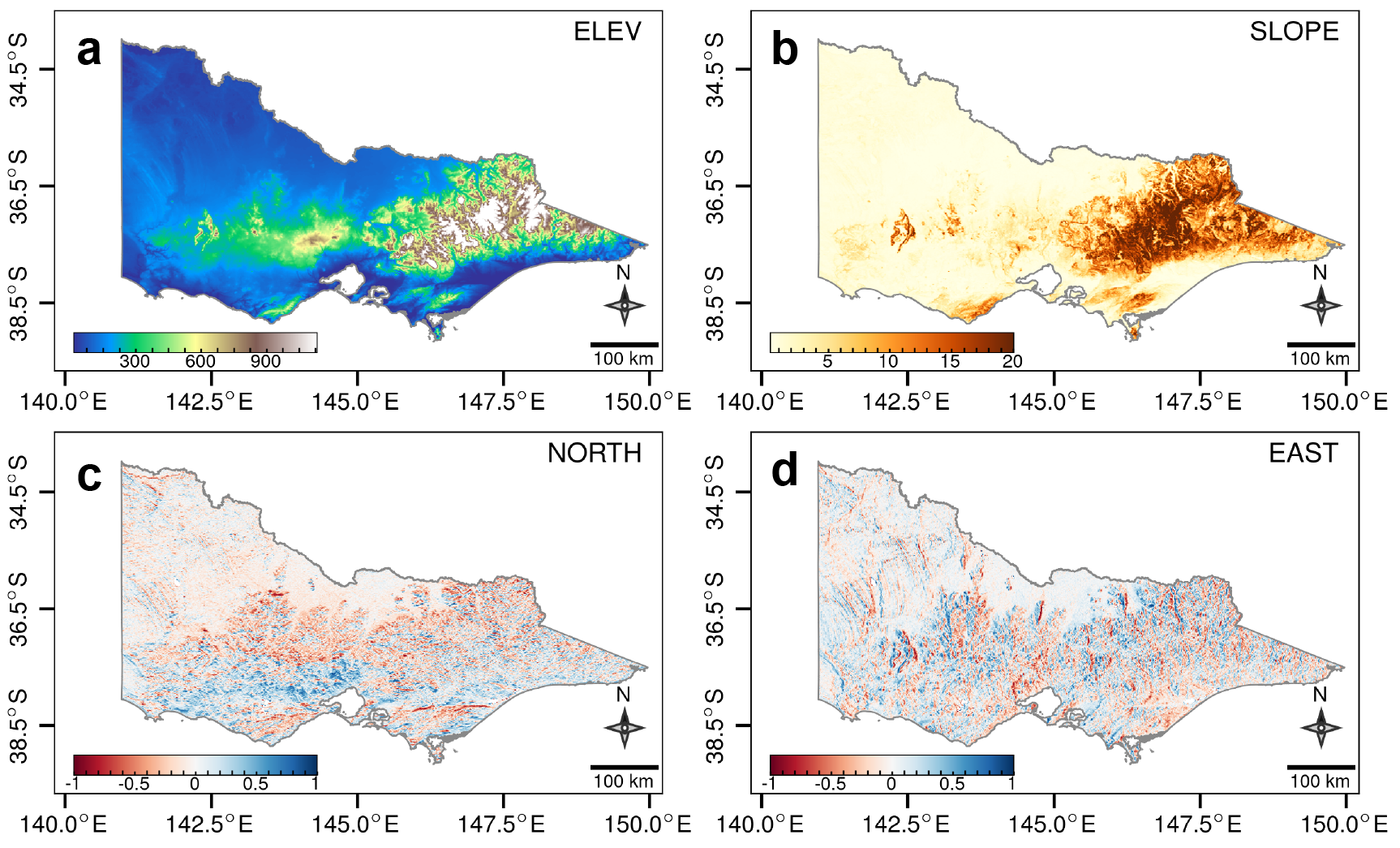}
	\caption{Spatial distribution of topographic variables: elevation (ELEV),
		slope (SLOPE), northness (NORTH), and eastness (EAST).}
	\label{fig:topographic}
\end{figure}
% -------------------------------------------------------

\subsection{Land Cover (Fig.~\ref{fig:landcover})}\label{subsec:lc}
\vspace{0.1cm}

\noindent \textbf{5--9 -- Land Cover Type} ([LC\_CRP, LC\_FRS, LC\_GRS, LC\_SVN, LC\_SHR]):
\begin{itemize}
	\item \textit{Unit:} NA, binary variables
	\item \textit{Calculation:} The \textit{mode} IGBP land cover class \citep{lovelandInternationalGeosphereBiosphere1997} within each 1~km cell was identified. We merged some of the original classes: Cropland (classes~12 \& 14), Forest (classes~1--5), Grassland (class~10), Savanna (classes~8--9), and Shrubland (classes~6--7). Remaining classes (urban areas, snow, waterbody, etc.) were merged into a single category to avoid high cardinality. The resulting grouped class was then converted into five binary indicators, with the remaining land-cover types serving as the reference category.
	\item \textit{Temporal aggregation:} Mode class across 2017--2025. As 2025 data were not yet available at the time of writing, the 2024 land cover map was used in place of 2025.
\end{itemize}

\begin{figure}[htbp]
	\centering
	\includegraphics[width=0.5\linewidth]{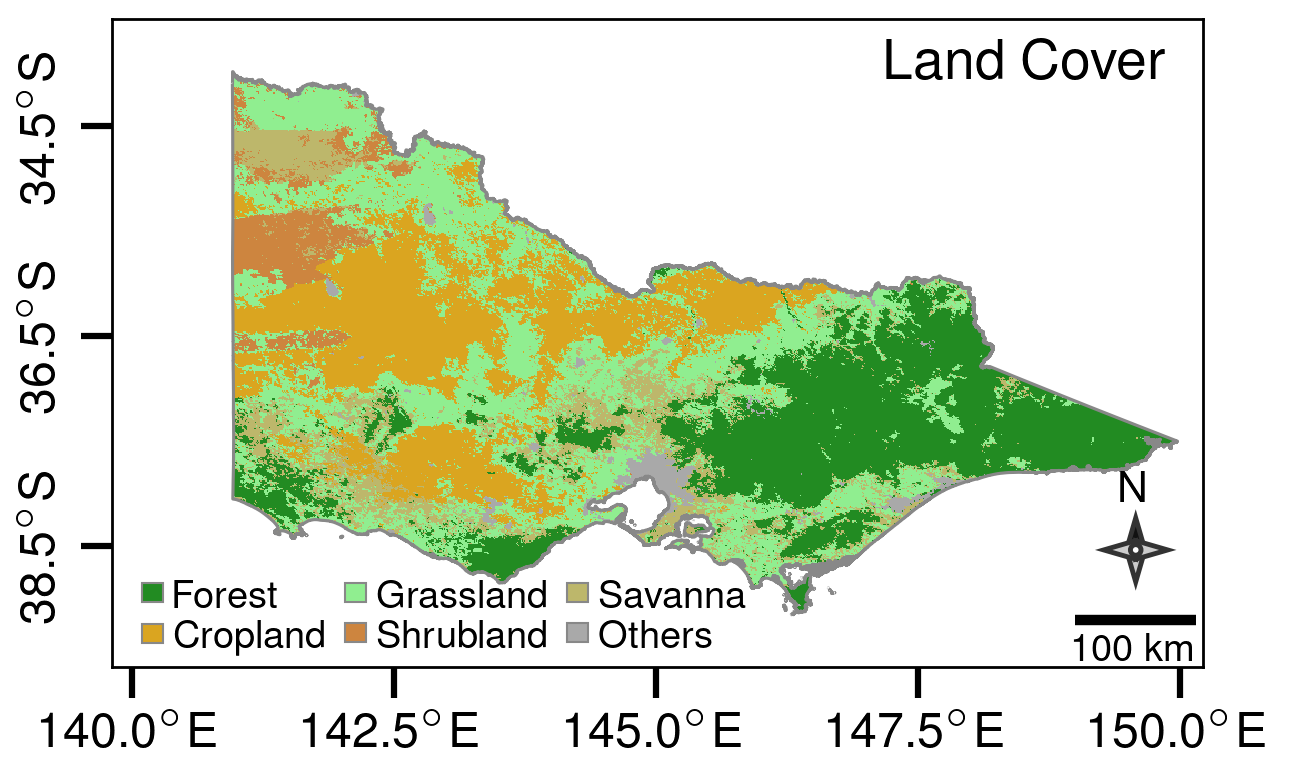}
	\caption{Land cover classes used in this study (merged from IGBP classification)}
	\label{fig:landcover}
\end{figure}
\vspace{0.1cm}
% -------------------------------------------------------
\subsection{Precipitation (Fig.~\ref{fig:precipitation})}\label{subsec:precip}
\vspace{0.1cm}
\noindent \textbf{10 -- Total Annual Precipitation} (PREC):
\begin{itemize}
	\item \textit{Unit:} mm
	\item \textit{Calculation:} Sum of daily rainfall within each calendar year $t$:
	\begin{equation}
		\text{PREC}_t = \sum\nolimits_{d=1}^{N_t} r_d
	\end{equation}
	where $r_d$ is daily rainfall (mm) on day $d$ and $N_t$ is the number of days in year $t$.
	\item \textit{Temporal aggregation:} Mean of annual totals across 2017--2025.
\end{itemize}

\noindent \textbf{11 -- Max Consecutive Dry Days in Fire Season} (CDD):
\begin{itemize}
	\item \textit{Unit:} days
	\item \textit{Yearly dataset calculation:} A day is \textit{dry} if daily rainfall $< 1.0$~mm. Within each calendar year, the length of the longest consecutive dry-day streak is identified separately for the two Australian fire-season windows ($\mathcal{W}_1 =$ Jan--Apr and $\mathcal{W}_2 =$ Oct--Dec). The annual value is the maximum of the two.
	\item \textit{Cross-sectional dataset calculation:} Daily rainfall records for 2017--2025 are treated as a single long time series. The length of the longest consecutive dry-day streak is computed for each of the following ten fire-season windows, which span calendar-year boundaries to reflect the continuity of the Australian fire season:
	\begin{itemize}
		\item Jan 2017 -- Apr 2017
		\item Oct 2017 -- Apr 2018
		\item Oct 2018 -- Apr 2019
		\item $\dots$
		\item Oct 2023 -- Apr 2024
		\item Oct 2024 -- Apr 2025
		\item Oct 2025 -- Dec 2025
	\end{itemize}
	The cross-sectional summary is then the maximum streak length across all windows:
	\begin{equation}
		\mathrm{CDD}^{\mathrm{summary}} = \max_{s \in \mathcal{S}}\, L_{s}
	\end{equation}
	where $\mathcal{S}$ denotes the set of fire-season windows listed above and $L_s$ denotes the length (in days) of the longest consecutive dry-day streak within window $s$. Note that this summary is computed directly from the full daily time series and is \emph{not} equivalent to the maximum of the annual values.
\end{itemize}

\noindent \textbf{12 -- Precipitation Seasonality} (PRECS):
\begin{itemize}
	\item \textit{Unit:} NA
	\item \textit{Calculation:} Coefficient of variation of 12 monthly precipitation totals:
	\begin{equation}
		\text{PRECS}_t = 100 \times \frac{s_{P,t}}{\bar{P}_t},
		\qquad s_{P,t} = \sqrt{\frac{1}{11}\sum_{k=1}^{12}(P_{k,t}-\bar{P}_t)^2}
	\end{equation}
	where $P_{k,t}$ is the total precipitation in month $k$ of year $t$, and $\bar{P}_t$ is the mean monthly precipitation in year $t$.
\end{itemize}

\begin{figure}[htbp]
	\centering
	\includegraphics[width=\linewidth]{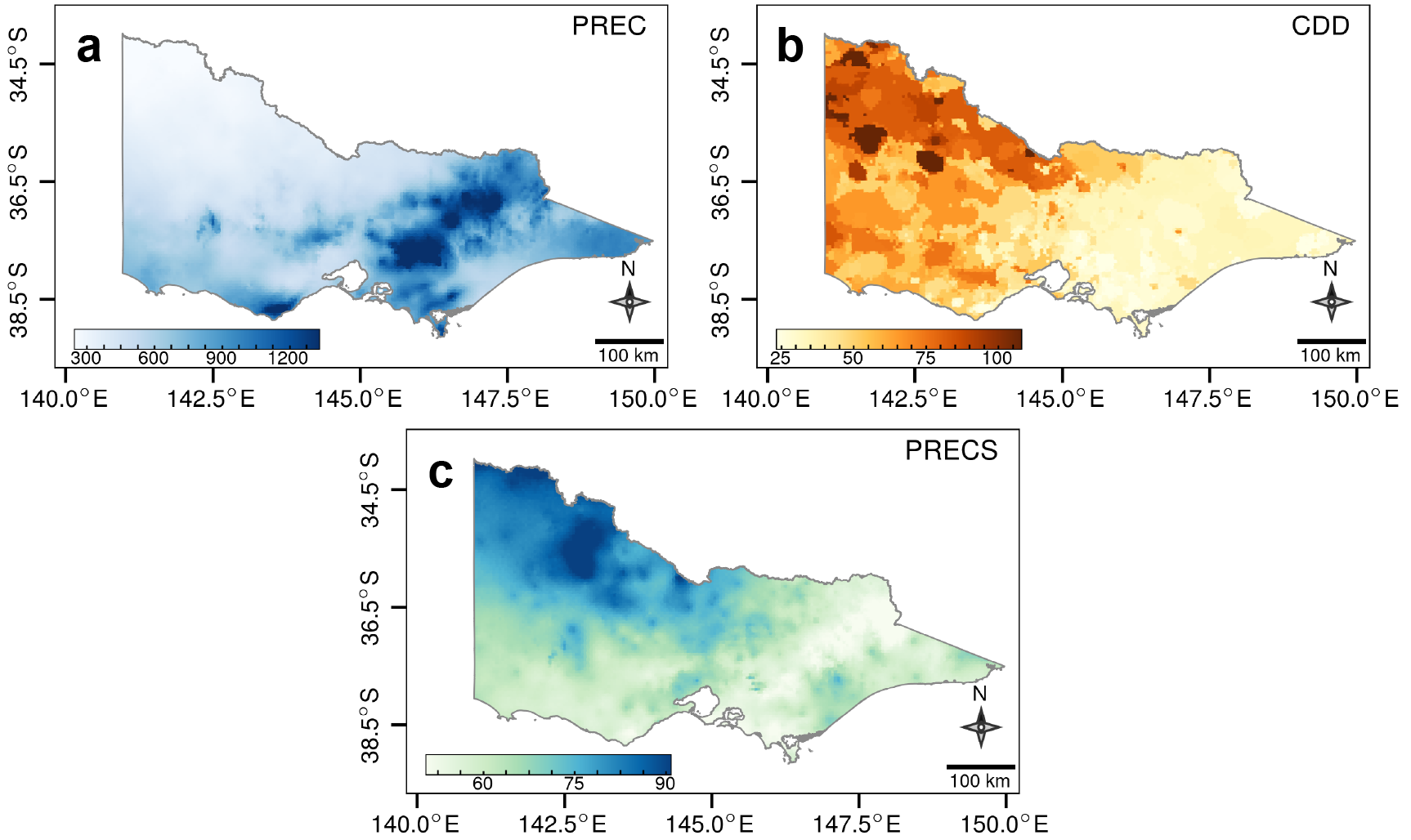}
	\caption{Spatial distribution of precipitation variables: total annual
		precipitation (PREC), maximum consecutive dry days during fire
		season windows (CDD), and precipitation seasonality (PRECS).}
	\label{fig:precipitation}
\end{figure}
% -------------------------------------------------------
\subsection{Temperature and Radiation (Fig.~\ref{fig:temperature})}\label{subsec:temp}
\vspace{0.1cm}
\noindent All temperature variables are derived from SILO daily $T_{\max,d}$ and $T_{\min,d}$, where $d$ indexes days. For notational brevity, year and month subscripts are omitted from $T_{\max,d}$ and $T_{\min,d}$. The summation range is implicit from the denominator: $N_t$ denotes the number of days in year $t$, and $N_{t,m}$ denotes the number of days in month $m$ of year $t$.

\vspace{0.1cm}
\noindent \textbf{13 -- Mean Temperature} (TMEAN):
\begin{itemize}
	\item \textit{Unit:} $^\circ$C
	\item \textit{Calculation:} Annual mean of daily mean temperature:
	\begin{equation}
		\text{TMEAN}_t = \frac{1}{N_t} \sum_{d} \frac{T_{\max,d} + T_{\min,d}}{2}
	\end{equation}
	\item \textit{Temporal aggregation:} Mean of $\text{TMEAN}_t$ across years.
\end{itemize}

\noindent \textbf{14 -- Maximum Temperature} (TMAX):
\begin{itemize}
	\item \textit{Unit:} $^\circ$C
	\item \textit{Calculation:} Annual maximum of daily $T_{\max,d}$:
	\begin{equation}
		\text{TMAX}_t = \max_{d} T_{\max,d}
	\end{equation}
	\item \textit{Temporal aggregation:} Maximum of $\text{TMAX}_t$ across years.
\end{itemize}

\noindent \textbf{15 -- Mean Monthly Temperature Range} (MTR):
\begin{itemize}
	\item \textit{Unit:} $^\circ$C
	\item \textit{Calculation:} The mean diurnal temperature range within month $m$ of year $t$ is:
	\begin{equation}
		\text{DTR}_{t,m} = \frac{1}{N_{t,m}} \sum_{d} \left( T_{\max,d} - T_{\min,d} \right)
	\end{equation}
	MTR for year $t$ is then the mean across all 12 months:
	\begin{equation}
		\text{MTR}_t = \frac{1}{12} \sum_{m=1}^{12} \text{DTR}_{t,m}
	\end{equation}
	\item \textit{Temporal aggregation:} Mean of $\text{MTR}_t$ across years.
\end{itemize}

\noindent \textbf{16 -- Temperature Annual Range} (TAR):
\begin{itemize}
	\item \textit{Unit:} $^\circ$C
	\item \textit{Calculation:} Difference between the annual maximum and annual minimum daily temperature:
	\begin{equation}
		\text{TAR}_t = \max_{d} T_{\max,d} - \min_{d} T_{\min,d}
	\end{equation}
	\item \textit{Temporal aggregation:} Mean of $\text{TAR}_t$ across years.
\end{itemize}

\noindent \textbf{17 -- Temperature Seasonality} (TS):
\begin{itemize}
	\item \textit{Unit:} NA
	\item \textit{Calculation:} Sample standard deviation of 12 monthly mean temperatures within year $t$, scaled by 100. The monthly mean temperature is:
	\begin{equation}
		\bar{T}_{t,m} = \frac{1}{N_{t,m}} \sum_{d} \frac{T_{\max,d} + T_{\min,d}}{2}
	\end{equation}
	Then:
	\begin{equation}
		\text{TS}_t = 100 \times \sqrt{\frac{1}{11} \sum_{m=1}^{12} \left( \bar{T}_{t,m} - \frac{1}{12}\sum_{m=1}^{12}\bar{T}_{t,m} \right)^2}
	\end{equation}
	\item \textit{Temporal aggregation:} Mean of $\text{TS}_t$ across years.
\end{itemize}

\noindent \textbf{18 -- Isothermality} (ISO):
\begin{itemize}
	\item \textit{Unit:} \%
	\item \textit{Calculation:} Ratio of mean monthly temperature range (variable 15) to annual temperature range (variable 16), expressed as a percentage:
	\begin{equation}
		\text{ISO}_t = 100 \times \frac{\text{MTR}_t}{\text{TAR}_t}
	\end{equation}
	\item \textit{Temporal aggregation:} Mean of $\text{ISO}_t$ across years.
\end{itemize}

\noindent \textbf{19 -- Max Consecutive Hot Days in Fire Season} (CHD):
\begin{itemize}
	\item \textit{Unit:} days
	\item \textit{Yearly dataset calculation:} Let $\tau_{90}$ denote the cell-level 90th percentile of daily $T_{\max,d}$ over all days in 2017--2025. A day $d$ is \textit{hot} if $T_{\max,d} > \tau_{90}$. Within each calendar year $t$, the length of the longest consecutive hot-day streak is identified separately for $\mathcal{W}_1 =$ Jan--Apr and $\mathcal{W}_2 =$ Oct--Dec. The annual value is the maximum of the two.
	\item \textit{Cross-sectional dataset calculation:} The same threshold $\tau_{90}$ is applied to the full daily $T_{\max,d}$ time series for 2017--2025. The length of the longest consecutive hot-day streak is computed for each of the ten fire-season windows defined in CDD (variable 11), and the cross-sectional summary is the maximum streak length across all windows. As with CDD, this is computed directly from the full time series and is \emph{not} the maximum of the annual values.
\end{itemize}

\noindent \textbf{20 -- Mean Solar Radiation} (SR):
\begin{itemize}
	\item \textit{Unit:} MJ~m$^{-2}$~day$^{-1}$
	\item \textit{Calculation:} Let $R_d$ denote daily solar radiation on day $d$. Annual mean:
	\begin{equation}
		\text{SR}_t = \frac{1}{N_t} \sum_{d} R_d
	\end{equation}
	\item \textit{Temporal aggregation:} Mean of $\text{SR}_t$ across years.
\end{itemize}
\begin{figure}[htbp]
	\centering
	\includegraphics[width=\linewidth]{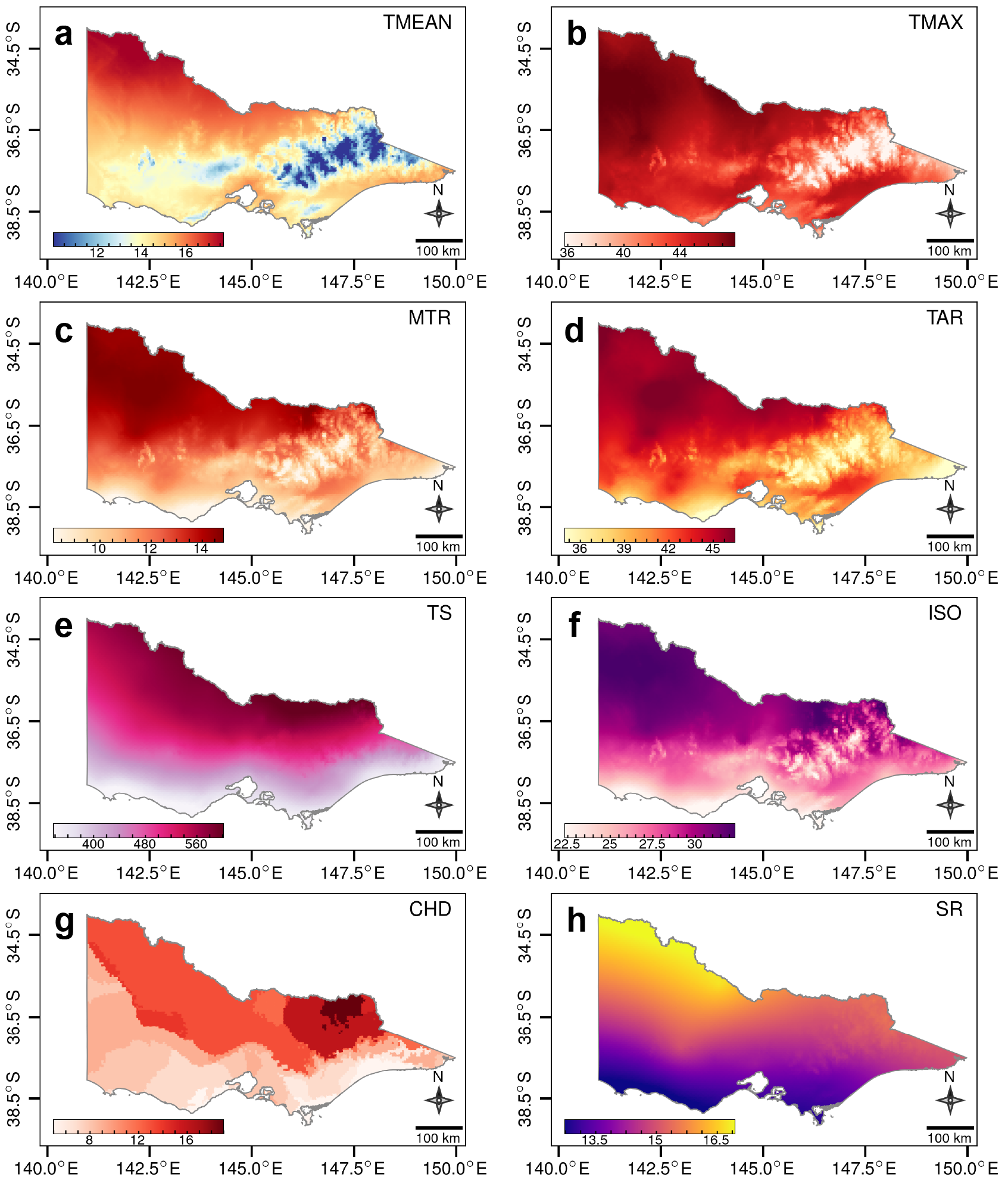}
	\caption{Temperature and radiation variables, including
		mean temperature (TMEAN), maximum temperature (TMAX),
		mean monthly temperature range (MTR), temperature annual
		range (TAR), temperature seasonality (TS), isothermality
		(ISO), maximum consecutive hot days in fire season (CHD),
		and mean solar radiation (SR).}
	\label{fig:temperature}
\end{figure}

% -------------------------------------------------------
\subsection{Wind (Fig.~\ref{fig:wind})}\label{subsec:wind}
\vspace{0.1cm}
\noindent \textbf{21 -- Mean Wind Speed} (WIND):

\begin{itemize}
	\item \textit{Unit:} m~s$^{-1}$
	\item \textit{Calculation:} Annual mean wind speed is computed as the arithmetic mean of the 12 monthly wind-speed values within each calendar year.
	\item \textit{Temporal aggregation:} Mean across years.
\end{itemize}

\begin{figure}[htbp]
	\centering
	\includegraphics[width=0.5\linewidth]{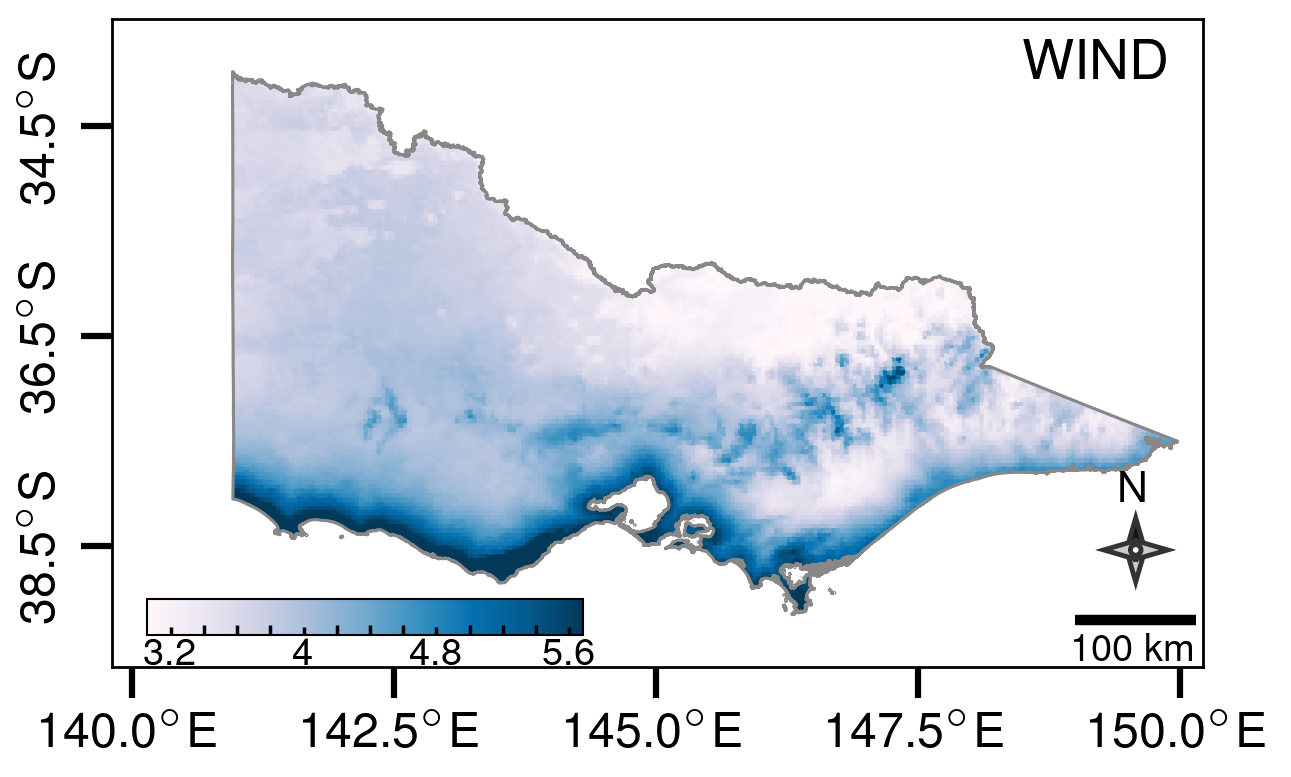}
	\caption{Spatial distribution of mean wind speed (WIND).}
	\label{fig:wind}
\end{figure}
% -------------------------------------------------------

\subsection{Vegetation Indices (Fig.~\ref{fig:vegetation})}\label{subsec:veg}
\vspace{0.1cm}

\noindent \textbf{22 -- Mean NDVI} (NDVI) and \textbf{23 -- Maximum NDVI} (NDVIMAX):
\begin{itemize}
	\item \textit{Unit:} NA
	\item \textit{Calculation:} Annual mean and maximum NDVI are computed from 16-day composites and averaged to 1~km.
	\item \textit{Temporal aggregation:} Mean across years (NDVI). Maximum across years (NDVIMAX).
\end{itemize}

\noindent \textbf{24 -- Mean LAI} (LAI) and \textbf{25 -- Maximum LAI} (LAIMAX):
\begin{itemize}
	\item \textit{Unit:} m$^2$~m$^{-2}$
	\item \textit{Calculation:} Annual mean and maximum are computed from 8-day composites.
	\item \textit{Temporal aggregation:} Mean across years (LAI). Maximum across years (LAIMAX).
\end{itemize}

\begin{figure}[htbp]
	\centering
	\includegraphics[width=\linewidth]{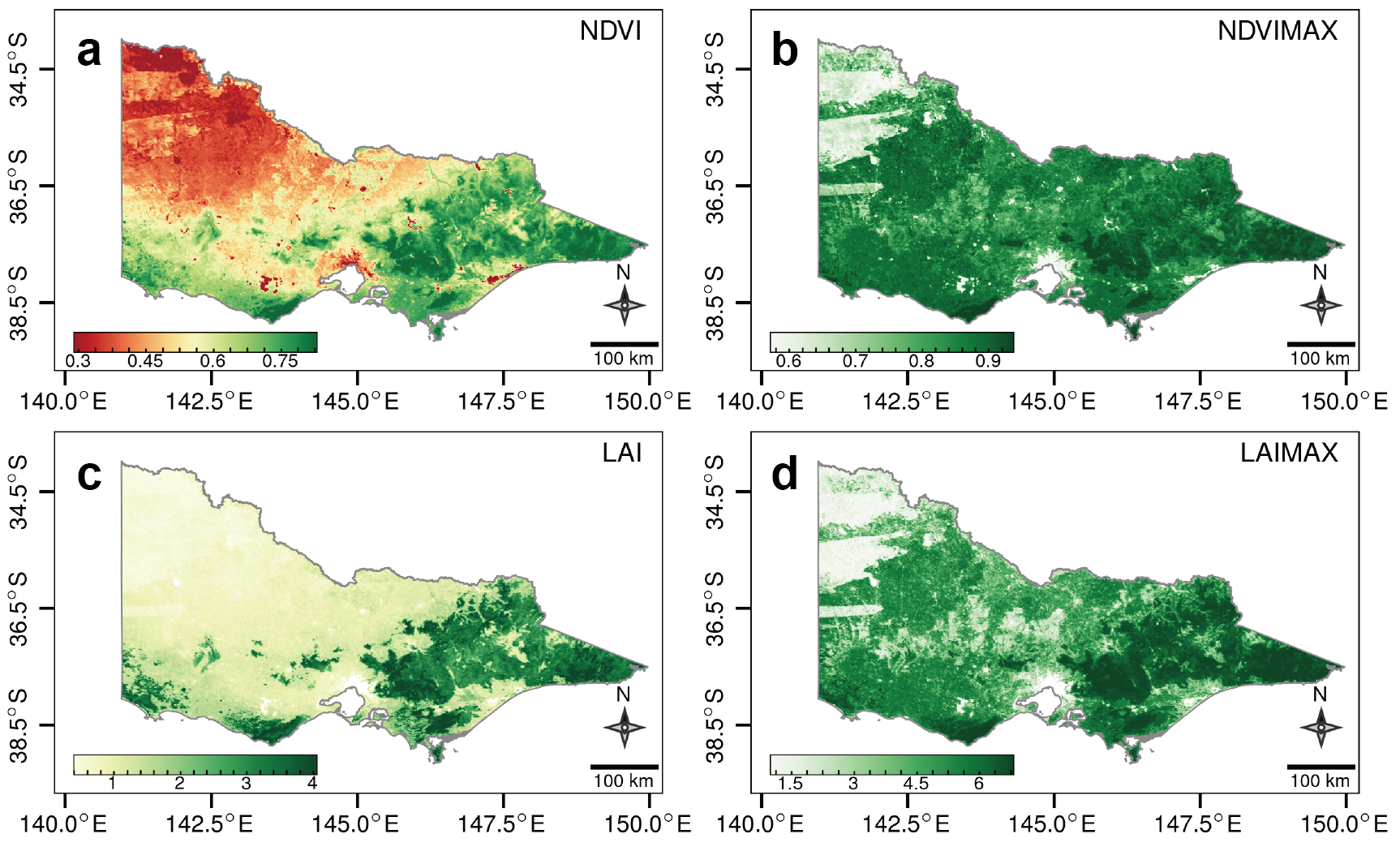}
	\caption{Vegetation index variables: mean NDVI (NDVI),
		maximum NDVI (NDVIMAX), mean leaf area index (LAI), and maximum
		leaf area index (LAIMAX).}
	\label{fig:vegetation}
\end{figure}

% -------------------------------------------------------
\subsection{Fire Weather Indices (Fig.~\ref{fig:fireweather})}\label{subsec:ffdi}
\vspace{0.1cm}
\noindent \textbf{26 -- Mean FFDI} (FFDI):
\begin{itemize}
	\item \textit{Unit:} NA
	\item \textit{Calculation:} Calendar-year mean of daily FFDI.
	\item \textit{Temporal aggregation:} Mean across years.
\end{itemize}

\noindent \textbf{27 -- Max Consecutive FFDI $>$ P90 Days in Fire Season} (CFFDI):
\begin{itemize}
	\item \textit{Unit:} days
	\item \textit{Yearly dataset calculation:} A day is \textit{high-FFDI} if daily FFDI exceeds the cell-level 90th percentile of daily FFDI over 2017--2025. Within each calendar year, the length of the longest consecutive high-FFDI streak is identified separately for $\mathcal{W}_1 =$ Jan--Apr and $\mathcal{W}_2 =$ Oct--Dec. The annual value is the maximum of the two.
	\item \textit{Cross-sectional dataset calculation:} The same high-FFDI threshold is applied to the full daily FFDI time series for 2017--2025. The length of the longest consecutive high-FFDI streak is computed for each of the ten fire-season windows defined in CDD (variable 11), and the cross-sectional summary is the maximum streak length across all windows.
\end{itemize}

\noindent \textbf{28 -- Number of Days FFDI $>$ 50 in Fire Season} (FFDI50):
\begin{itemize}
	\item \textit{Unit:} days
	\item \textit{Yearly dataset calculation:} Count of days within each calendar year's fire-season windows ($\mathcal{W}_1 =$ Jan--Apr and $\mathcal{W}_2 =$ Oct--Dec) where daily FFDI $> 50$ (the ``Severe'' threshold \citep{bom2019ffdi}).
	\item \textit{Cross-sectional dataset calculation:} Count of days across all ten fire-season windows defined in CDD (variable 11) where daily FFDI $> 50$. The cross-sectional summary is the total count summed across all windows.
\end{itemize}

\begin{figure}[htbp]
	\centering
	\includegraphics[width=\linewidth]{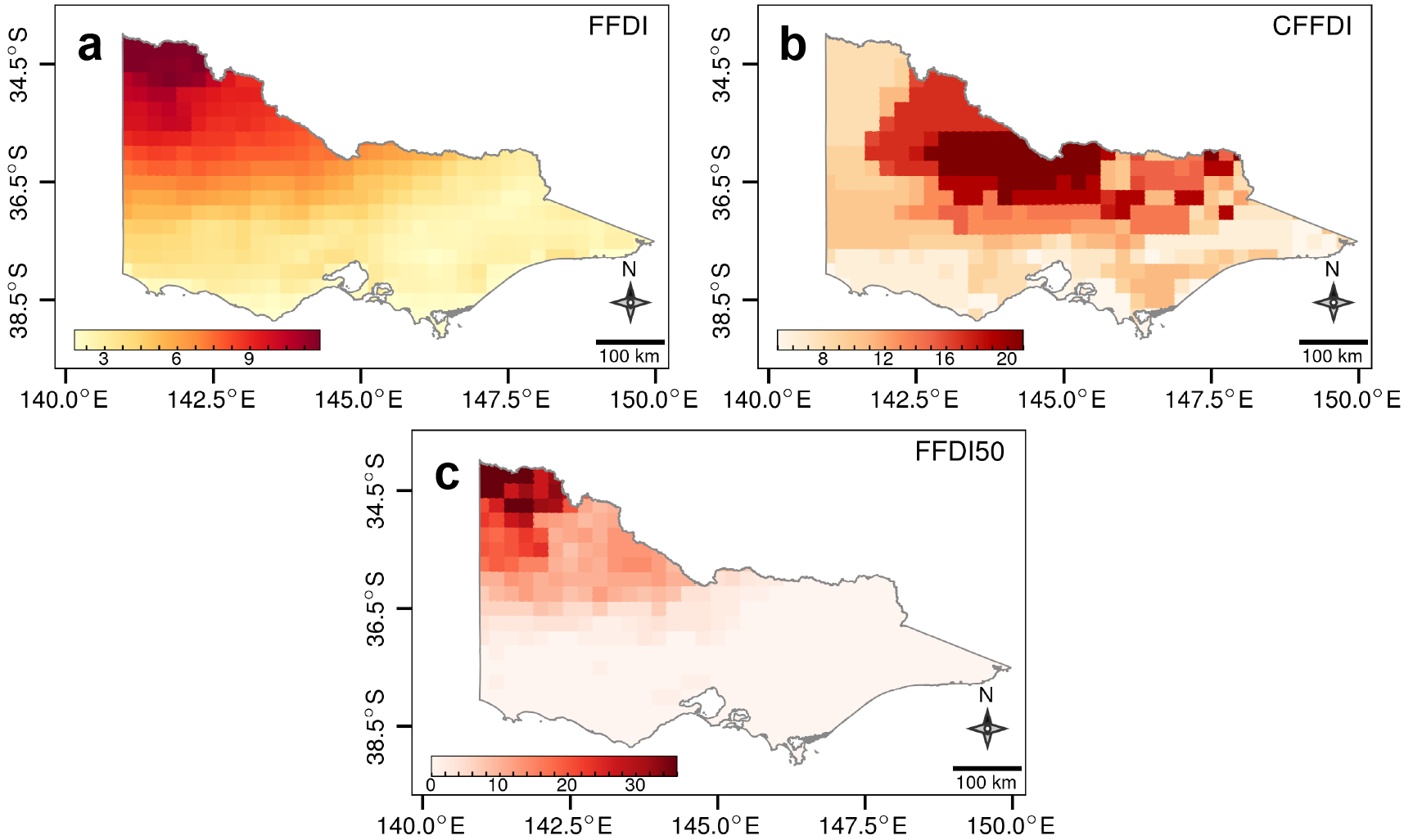}
	\caption{Spatial distribution of fire weather index variables: mean FFDI (FFDI), maximum consecutive high-FFDI days in fire season (CFFDI), and number of days exceeding FFDI threshold 50 in fire season (FFDI50).}
	\label{fig:fireweather}
\end{figure}

% -------------------------------------------------------
\subsection{Atmospheric Variables (Fig.~\ref{fig:atmospheric})}\label{subsec:atmos}
\vspace{0.1cm}
\noindent \textbf{29 -- Mean Vapour Pressure Deficit} (VPD):
\begin{itemize}
	\item \textit{Unit:} hPa
	\item \textit{Calculation:} Calendar-year mean of daily vapour pressure deficit.
	\item \textit{Temporal aggregation:} Mean across years.
\end{itemize}

\noindent \textbf{30 -- Mean Water Deficit} (WD):
\begin{itemize}
	\item \textit{Unit:} mm~day$^{-1}$
	\item \textit{Calculation:} Daily water deficit is Morton potential ET minus areal actual ET. The calendar-year mean is taken.
	\item \textit{Temporal aggregation:} Mean across years.
\end{itemize}

\begin{figure}[htbp]
	\centering
	\includegraphics[width=\linewidth]{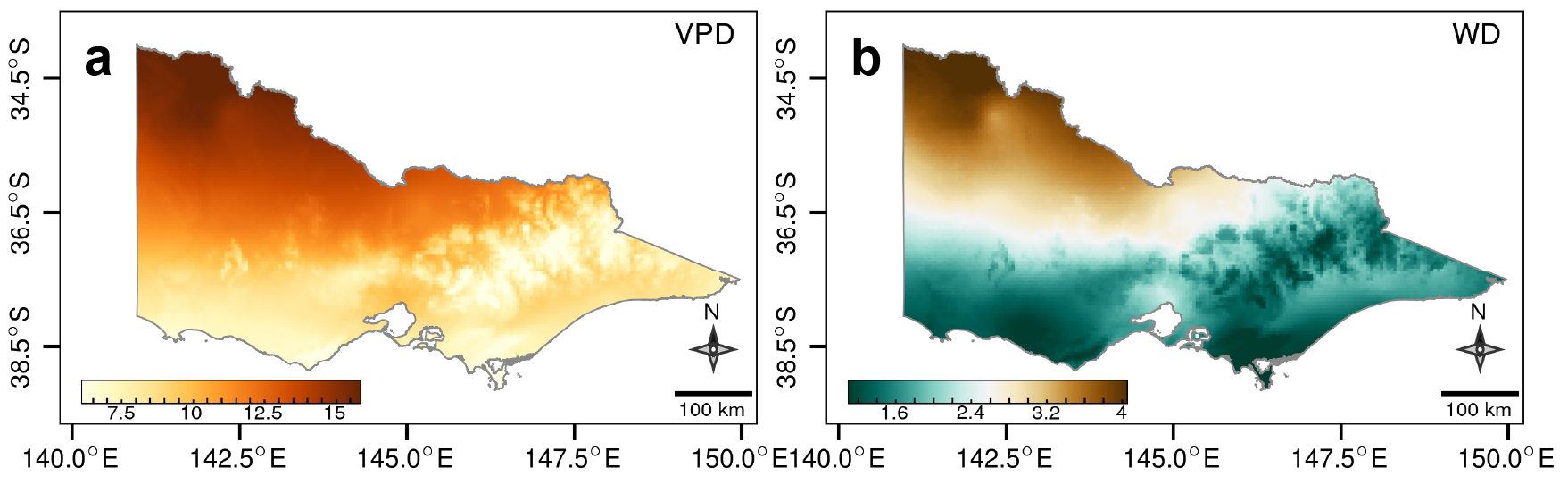}
	\caption{Atmospheric moisture stress variables: vapour pressure deficit (VPD) and water deficit (WD).}
	\label{fig:atmospheric}
\end{figure}
% -------------------------------------------------------
\subsection{Proximity Variables (Fig.~\ref{fig:proximity})}\label{subsec:prox}
\vspace{0.1cm}
\noindent All proximity variables are static (time-invariant) and computed directly on the 1~km reference grid.

\vspace{0.1cm}
\noindent \textbf{31 -- Distance to Major Road} (DROAD):
\begin{itemize}
	\item \textit{Unit:} m
	\item \textit{Calculation:} Road geometries (motorway, trunk, primary, secondary, and their link variants) are retrieved. A Euclidean Distance Transform yields per-cell distance to the nearest qualified road.
\end{itemize}

\noindent \textbf{32 -- Distance to Waterbody} (DWATER):
\begin{itemize}
	\item \textit{Unit:} m
	\item \textit{Calculation:} Waterbody polygons are extracted from the source dataset. A Euclidean Distance Transform yields per-cell distance to the nearest waterbody.
\end{itemize}

\begin{figure}[htbp]
	\centering
	\includegraphics[width=\linewidth]{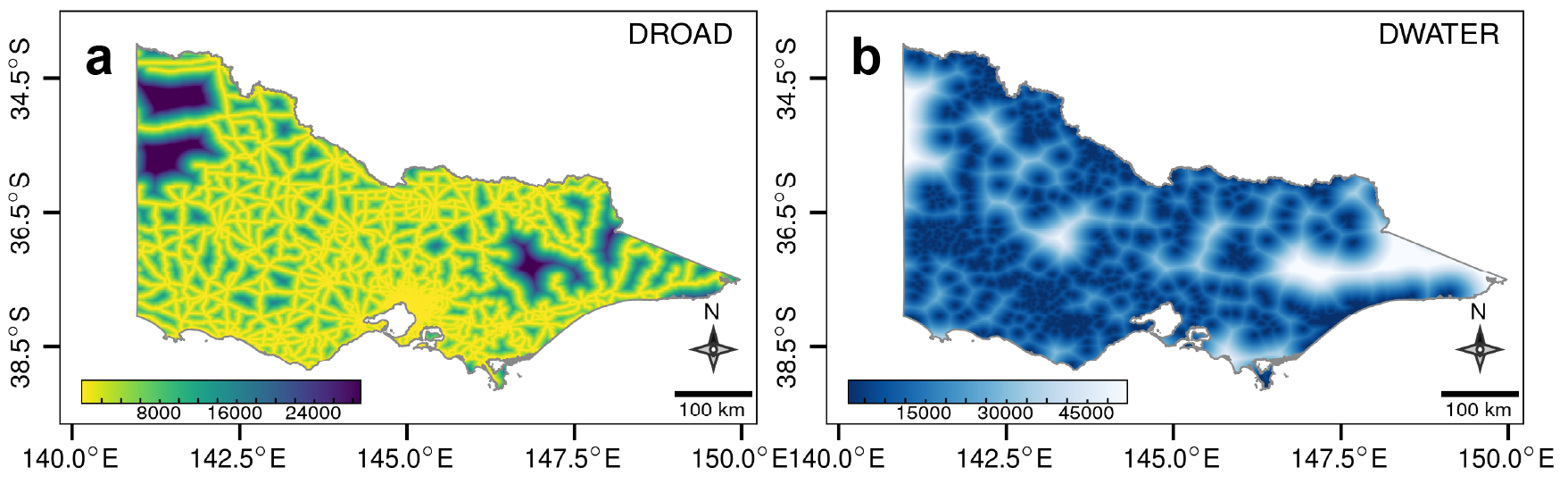}
	\caption{Proximity variables: distance to major road (DROAD) and distance to waterbody (DWATER).}
	\label{fig:proximity}
\end{figure}

% -------------------------------------------------------
\subsection{Soil Moisture (Fig.~\ref{fig:soil-moisture})}\label{subsec:soil}
\vspace{0.1cm}
\noindent Two products from the Australian Water Outlook (AWO) AWRA-L daily root-zone soil moisture were used: absolute soil moisture (\texttt{sm\_pct}, fraction of root-zone fullness, 0--1) and relative soil moisture (daily percentile rank relative to the historical record, 0--1).

\vspace{0.1cm}
\noindent \textbf{33 -- Mean Soil Moisture} (SM):
\begin{itemize}
	\item \textit{Unit:} NA
	\item \textit{Calculation:} Annual mean of Absolute \texttt{sm\_pct}.
	\item \textit{Temporal aggregation:} Mean across years.
\end{itemize}

\noindent \textbf{34 -- Minimum Soil Moisture} (SMMIN):
\begin{itemize}
	\item \textit{Unit:} NA
	\item \textit{Calculation:} Annual minimum of absolute \texttt{sm\_pct}, capturing the driest single day of the year.
	\item \textit{Temporal aggregation:} Minimum across years.
\end{itemize}

\noindent \textbf{35 -- Max Consecutive Dry Soil Days in Fire Season} (CDSD):
\begin{itemize}
	\item \textit{Unit:} days
	\item \textit{Yearly dataset calculation:} A day is defined as a \textit{dry-soil day} if relative soil moisture $< 0.10$ (below the 10th historical percentile). Within each calendar year, the length of the longest consecutive dry-soil streak is identified separately for $\mathcal{W}_1 =$ Jan--Apr and $\mathcal{W}_2 =$ Oct--Dec. The annual value is the maximum of the two.
	\item \textit{Cross-sectional dataset calculation:} The same dry-soil threshold is applied to the full daily relative soil moisture time series for 2017--2025. The length of the longest consecutive dry-soil streak is computed for each of the ten fire-season windows defined in CDD (variable 11), and the cross-sectional summary is the maximum streak length across all windows.
\end{itemize}

\begin{figure}[htbp]
	\centering
	\includegraphics[width=\linewidth]{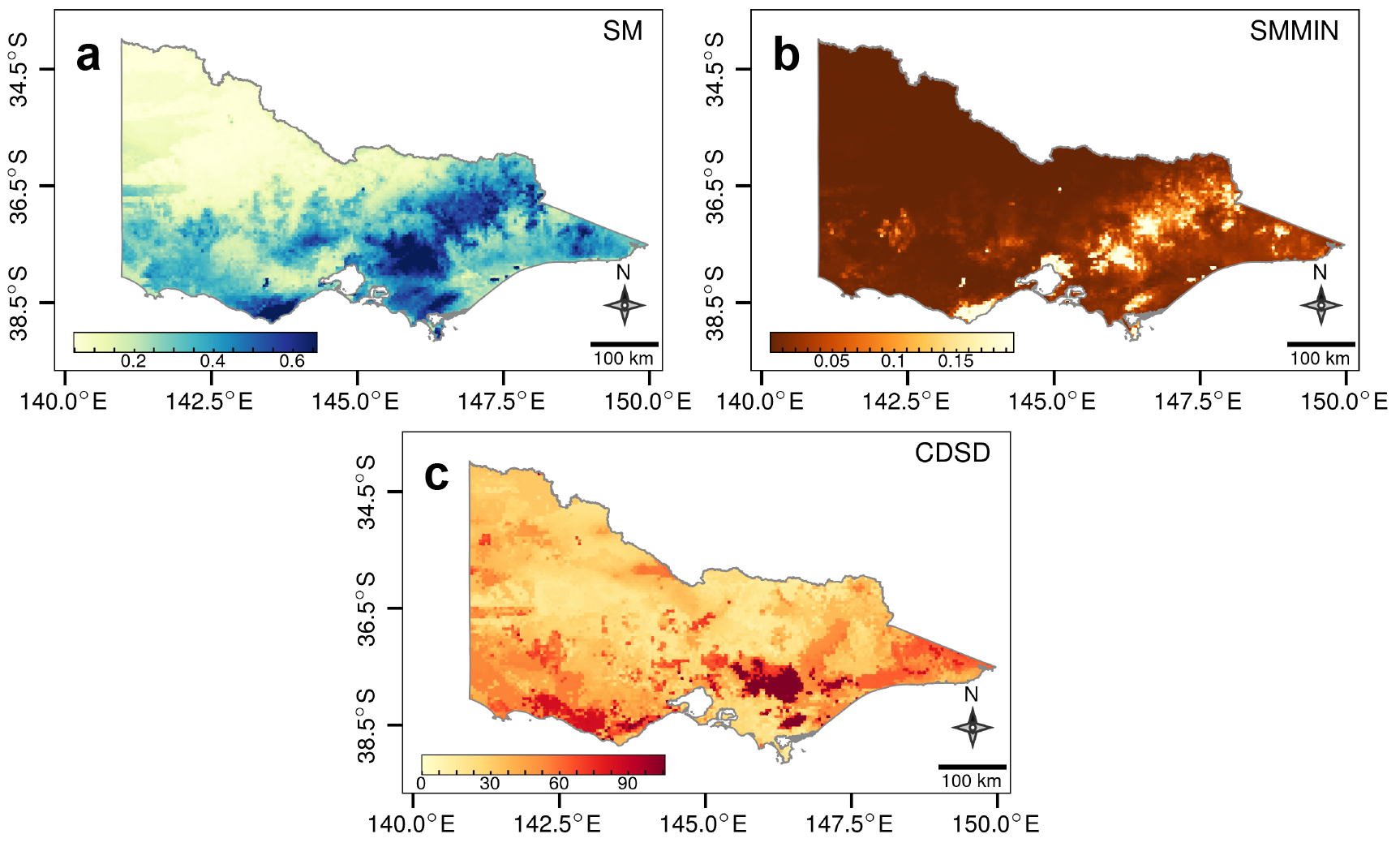}
	\caption{Soil moisture conditioning variables including Mean Soil Moisture (SM), Minimum Soil Moisture (SMMIN) and Max Consecutive Dry Soil Days in Fire Season (CDSD).}
	\label{fig:soil-moisture}
\end{figure}

\section{Spatial Autocorrelation Analysis}
\label{app:spatial_autocorrelation}

Spatial autocorrelation describes the tendency for nearby observations to have more similar values than distant observations \citep{websterGeostatisticsEnvironmentalScientists2007}. In wildfire susceptibility modelling, this is relevant to negative-sample selection: negatives located too close to fire-occurrence cells may be environmentally similar to positives but assigned opposite labels, while an overly large separation constraint can reduce the availability and representativeness of negative samples.

To inform the choice of positive-to-negative separation distance, we analysed the spatial autocorrelation structure of both the physical variables and the AlphaEarth Foundations (AEF) embedding dimensions. Spatial autocorrelation was quantified using empirical semivariograms, a standard geostatistical tool for characterising how attribute dissimilarity increases with spatial separation \citep{schabenbergerStatisticalMethodsSpatial2005}.
For a spatial variable $Z(\mathbf{s})$ observed at location $\mathbf{s}$, the empirical semivariogram at lag distance $h$ is defined as
\begin{equation}
	\hat{\gamma}(h)
	=
	\frac{1}{2\lvert N(h)\rvert}
	\sum_{(i,j)\in N(h)}
	\bigl[Z(\mathbf{s}_i)-Z(\mathbf{s}_j)\bigr]^{2},
\end{equation}
where $N(h)$ denotes the set of location pairs whose separation distance falls
within the lag bin centred at $h$, and $\lvert N(h)\rvert$ is the cardinality of that set. By definition, $\gamma(0)=0$, and $\hat{\gamma}(h)$ generally increases with $h$ as spatially distant locations tend to be less similar. Under the assumption of approximate second-order stationarity, the semivariogram is related to the covariance function by
\begin{equation}
	\gamma(h) = C(0) - C(h),
\end{equation}
where $C(0)$ is the marginal variance and $C(h)$ is the covariance between locations separated by distance $h$. Larger semivariance therefore indicates weaker spatial similarity.

Before semivariogram estimation, all variables were standardised using z-scores so that semivariance values were comparable across variables with different units and numerical ranges. For the physical dataset, four representative continuous variables were selected to cover major environmental dimensions: slope, total precipitation, mean temperature, and maximum NDVI. For the AEF embedding dataset, semivariograms were computed separately for all 64 embedding dimensions. The mean semivariogram across dimensions was then used to summarise the overall spatial dependence structure of the embedding space. For both datasets, 100,000 valid pixels were randomly sampled, and empirical semivariograms were estimated over lag distances from 0 to 20~km.
\begin{figure}[htbp]
	\centering
	\includegraphics[width=\linewidth]{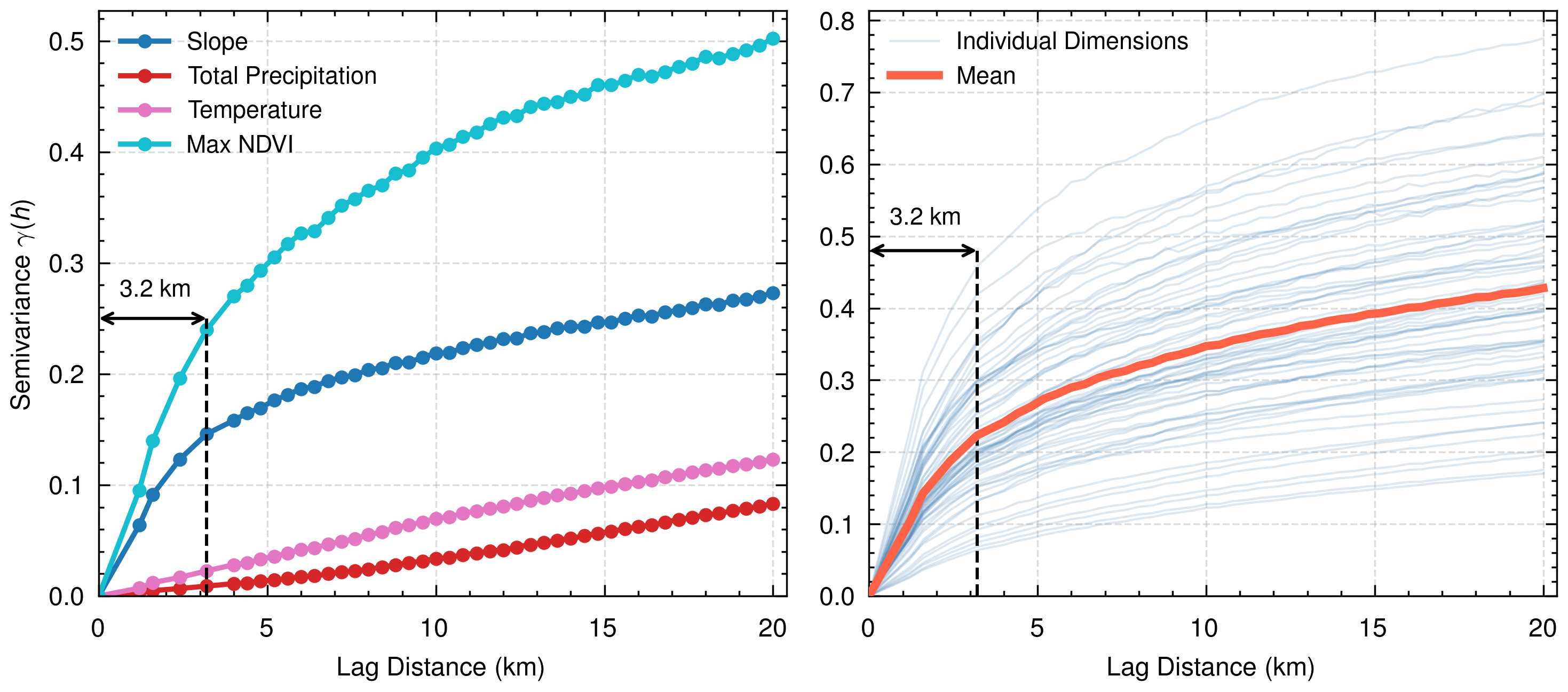}
	\caption{Empirical semivariograms for representative physical wildfire driving factors (left) and AEF embedding dimensions (right).}
	\label{fig:semivariogram}
\end{figure}

Fig.~\ref{fig:semivariogram} reveals clear short-range spatial autocorrelation in both the physical variables and the AEF embeddings, although the strength and decay rate of this dependence differ across variables. At the 3.2~km reference distance, slope, maximum NDVI, and the mean AEF semivariogram have already shown a marked increase from the origin, indicating that a substantial portion of the strongest local similarity has been reduced by this distance. Accordingly, a minimum positive-to-negative separation distance of 3~km was adopted in the main sampling design --- close to this reference distance and
providing a practical balance between reducing label ambiguity and preserving the availability of negative-sample candidates.

\section{Hyperparameter Search Spaces and Best Configurations}\label{app:hyperparams}

Tables~\ref{tab:hp_tree}--\ref{tab:hp_temporal} report the hyperparameter search spaces and optimal configurations selected by Optuna for each model and feature representation. Ranges annotated with ``log-uniform'' indicate sampling on a logarithmic scale over the specified interval.

\clearpage

% Keep the three compact hyperparameter tables together on one page.
\thispagestyle{supplementary}
\begingroup
\scriptsize
\renewcommand{\arraystretch}{0.88}
\setlength{\intextsep}{5pt}
\setlength{\textfloatsep}{5pt}
\setlength{\abovecaptionskip}{2pt}
\setlength{\belowcaptionskip}{2pt}
\null\vfill

\begin{table}[H]
	\centering
	\caption{Hyperparameter search spaces and best configurations for 
		cross-sectional wildfire susceptibility models. The MLP architecture ([128, 64] hidden 
		units) and weight decay ($10^{-4}$) are fixed across all 
		settings.}\label{tab:hp_tree}
	\setlength{\tabcolsep}{4pt}
	\begin{tabular}{@{} l l l l l @{}}
		\toprule
		\textbf{Model} & \textbf{Parameter} & \textbf{Search Range / Values} & \textbf{Best (Physical)} & \textbf{Best (Embedding)} \\
		\midrule
		\multirow{5}{*}{Random Forests}
		& \texttt{n\_estimators} & $\{300, 400, \dots, 1000\}$ & 800 & 800 \\
		& \texttt{max\_depth} & $\{6, 7, \dots, 25\}$ & 25 & 23 \\
		& \texttt{min\_samples\_leaf} & $\{3, 4, \dots, 30\}$ & 3 & 3 \\
		& \texttt{max\_features} & $\{\text{sqrt}, \text{log2}, 0.3, 0.5, 0.7\}$ & sqrt & sqrt \\
		& \texttt{ccp\_alpha} & $[10^{-6}, 10^{-2}]$, log-uniform & $1.96 \times 10^{-5}$ & $2.87 \times 10^{-6}$ \\
		\midrule
		\multirow{9}{*}{XGBoost}
		& \texttt{learning\_rate} & $[0.01, 0.15]$, log-uniform & 0.131 & 0.049 \\
		& \texttt{max\_depth} & $\{3, 4, \dots, 8\}$ & 5 & 7 \\
		& \texttt{subsample} & $[0.6, 1.0]$ & 0.810 & 0.657 \\
		& \texttt{colsample\_bytree} & $[0.5, 1.0]$ & 0.534 & 0.663 \\
		& \texttt{reg\_lambda} & $[0.1, 20]$, log-uniform & 2.28 & 4.85 \\
		& \texttt{reg\_alpha} & $[10^{-4}, 5]$, log-uniform & 0.113 & 0.629 \\
		& \texttt{min\_child\_weight} & $[2.0, 50.0]$ & 2.15 & 6.53 \\
		& \texttt{gamma} & $[0.0, 5.0]$ & 0.33 & 0.77 \\
		& \texttt{n\_estimators} & $\{300, 350, \dots, 3000\}$ & 2900 & 3000 \\
		\midrule
		\multirow{8}{*}{LightGBM}
		& \texttt{learning\_rate} & $[0.01, 0.15]$, log-uniform & 0.031 & 0.012 \\
		& \texttt{num\_leaves} & $\{16, 17, \dots, 128\}$ & 72 & 91 \\
		& \texttt{min\_data\_in\_leaf} & $\{20, 21, \dots, 300\}$ & 40 & 34 \\
		& \texttt{feature\_fraction} & $[0.5, 1.0]$ & 0.879 & 0.801 \\
		& \texttt{bagging\_fraction} & $[0.6, 1.0]$ & 0.966 & 0.706 \\
		& \texttt{lambda\_l2} & $[0.1, 30]$, log-uniform & 0.52 & 1.34 \\
		& \texttt{lambda\_l1} & $[10^{-4}, 5]$, log-uniform & $1.11 \times 10^{-4}$ & $1.14 \times 10^{-4}$ \\
		& \texttt{n\_estimators} & $\{300, 350, \dots, 3000\}$ & 1450 & 2550 \\
		\midrule
		\multirow{4}{*}{MLP}
		& \texttt{dropout} & $\{0, 0.1, 0.2, 0.3, 0.4, 0.5\}$ & 0.1 & 0.2 \\
		& \texttt{lr} & $[10^{-5}, 10^{-2}]$, log-uniform, 10 values & $4.64 \times 10^{-3}$ & $2.15 \times 10^{-4}$ \\
		& \texttt{batch\_size} & $\{64, 128, 256\}$ & 256 & 64 \\
		& \texttt{max\_epochs} & $\{25, 50, 100, 200, 500\}$ & 200 & 200 \\
		\bottomrule
	\end{tabular}
\end{table}

\begin{table}[H]
	\centering
	\caption{Hyperparameter search spaces and best configurations for 
		CNN wildfire susceptibility models. The convolutional backbone (32 $\to$ 64 $\to$ 
		128 channels, $3 \times 3$ kernels), MLP head ([128, 64]), 
		epochs (50) and weight decay ($10^{-4}$) are fixed across all configurations.}\label{tab:hp_cnn_spatial}
	\setlength{\tabcolsep}{2.6pt}
	\footnotesize
	\begin{tabular}{@{} l l c c c c c c @{}}
		\toprule
		\multirow{2}{*}{\textbf{Parameter}} & \multirow{2}{*}{\textbf{Search Values}} & \multicolumn{3}{c}{\textbf{Best (Physical)}} & \multicolumn{3}{c}{\textbf{Best (Embedding)}} \\
		\cmidrule(lr){3-5} \cmidrule(lr){6-8}
		& & \textbf{CNN9} & \textbf{CNN17} & \textbf{CNN25} & \textbf{CNN9} & \textbf{CNN17} & \textbf{CNN25} \\
		\midrule
		\texttt{batch\_size} & $\{16, 32, 64\}$ & 32 & 32 & 32 & 64 & 64 & 32 \\
		\texttt{dropout}     & $\{0.2, 0.5, 0.8\}$ & 0.8 & 0.8 & 0.5 & 0.8 & 0.8 & 0.2 \\
		\texttt{lr}          & $[10^{-4}, 10^{-3}]$, log-uniform, 4 values & $4.64 \times 10^{-4}$ & $4.64 \times 10^{-4}$ & $4.64 \times 10^{-4}$ & $4.64 \times 10^{-4}$ & $4.64 \times 10^{-4}$ & $2.15 \times 10^{-4}$ \\
		\bottomrule
	\end{tabular}
\end{table}

\begin{table}[H]
	\centering
	\caption{Hyperparameter search spaces and best configurations for temporal models. All other hyperparameters follow the same fixed settings as the cross-sectional MLP and CNN models.}\label{tab:hp_temporal}
	\setlength{\tabcolsep}{0.4pt}
	\scriptsize
	\begin{tabular}{@{} l l c c c c c c c c @{}}
		\toprule
		\multirow{2}{*}{\textbf{Parameter}} & \multirow{2}{*}{\textbf{Search Values}} & \multicolumn{4}{c}{\textbf{Best (Physical)}} & \multicolumn{4}{c}{\textbf{Best (Embedding)}} \\
		\cmidrule(lr){3-6} \cmidrule(lr){7-10}
		& & \textbf{MLP} & \textbf{CNN9} & \textbf{CNN17} & \textbf{CNN25} & \textbf{MLP} & \textbf{CNN9} & \textbf{CNN17} & \textbf{CNN25} \\
		\midrule
		\multirow{2}{*}{\texttt{batch\_size}}
		& MLP: $\{64, 128, 256\}$
		& \multirow{2}{*}{64} & \multirow{2}{*}{64} & \multirow{2}{*}{32} & \multirow{2}{*}{64}
		& \multirow{2}{*}{64} & \multirow{2}{*}{64} & \multirow{2}{*}{64} & \multirow{2}{*}{64} \\
		& CNN: $\{16, 32, 64\}$ & & & & & & & & \\
		\midrule
		\multirow{2}{*}{\texttt{dropout}}
		& MLP: $\{0, 0.1, 0.2, 0.3, 0.4, 0.5\}$
		& \multirow{2}{*}{0.1} & \multirow{2}{*}{0.2} & \multirow{2}{*}{0.8} & \multirow{2}{*}{0.8}
		& \multirow{2}{*}{0.5} & \multirow{2}{*}{0.5} & \multirow{2}{*}{0.8} & \multirow{2}{*}{0.8} \\
		& CNN: $\{0.2, 0.5, 0.8\}$ & & & & & & & & \\
		\midrule
		\multirow{2}{*}{\texttt{lr}}
		& MLP: $[10^{-5}, 10^{-2}]$, log-uniform, 10 values
		& \multirow{2}{*}{$4.64{\times}10^{-3}$} & \multirow{2}{*}{$10^{-3}$} & \multirow{2}{*}{$4.64{\times}10^{-4}$} & \multirow{2}{*}{$4.64{\times}10^{-4}$}
		& \multirow{2}{*}{$4.64{\times}10^{-4}$} & \multirow{2}{*}{$4.64{\times}10^{-4}$} & \multirow{2}{*}{$4.64{\times}10^{-4}$} & \multirow{2}{*}{$4.64{\times}10^{-4}$} \\
		& CNN: $[10^{-4}, 10^{-3}]$, log-uniform, 
		 4 values & & & & & & & & \\
		\bottomrule
	\end{tabular}
\end{table}

\vfill\null
\endgroup
\clearpage

\section{Additional Results}
\label{app:additional_results}
\thispagestyle{supplementary}

This section provides additional figures and tables supporting the main results.

% Lead the section with Table S6 and Fig. S12 on the same page.  The local
% table-counter change preserves the published S4--S6 numbering while allowing
% the requested visual grouping; after this group the counter returns to S3.
\begingroup
\begin{samepage}
\begin{center}
	\refstepcounter{table}\label{tab:embedding_top_dims}
	\begin{minipage}{\linewidth}
		\normalsize\textbf{Table~\thetable}\par
		AEF embedding dimensions ranked among the top 15 features in tree-based cross-sectional models. A value of 1 marks dimensions appearing in the top 15 feature-importance ranking of the corresponding model, and 0 otherwise. The dimension index starts from 1.
	\end{minipage}
	\par\vspace{3pt}
	\normalsize
	\renewcommand{\arraystretch}{0.72}
	\setlength{\tabcolsep}{6pt}
	\begin{tabular}{@{} c c c c c @{}}
		\toprule
		\textbf{Dimension} & \textbf{LightGBM} & \textbf{XGBoost} & \textbf{Random Forest} & \textbf{Count} \\
		\midrule
		14 & 1 & 1 & 1 & 3 \\
		41 & 1 & 1 & 1 & 3 \\
		44 & 1 & 1 & 1 & 3 \\
		56 & 1 & 1 & 1 & 3 \\
		35 & 1 & 1 & 1 & 3 \\
		60 & 1 & 1 & 1 & 3 \\
		24 & 1 & 1 & 1 & 3 \\
		29 & 1 & 1 & 1 & 3 \\
		43 & 1 & 1 & 1 & 3 \\
		50 & 1 & 1 & 1 & 3 \\
		9  & 1 & 0 & 1 & 2 \\
		1  & 0 & 1 & 1 & 2 \\
		52 & 0 & 1 & 1 & 2 \\
		7  & 1 & 1 & 0 & 2 \\
		36 & 0 & 0 & 1 & 1 \\
		21 & 0 & 1 & 0 & 1 \\
		59 & 1 & 0 & 0 & 1 \\
		2  & 0 & 1 & 0 & 1 \\
		8  & 1 & 0 & 0 & 1 \\
		33 & 1 & 0 & 0 & 1 \\
		3  & 0 & 0 & 1 & 1 \\
		\bottomrule
	\end{tabular}
\end{center}

\vspace{-0.5em}
\begin{center}
	\includegraphics[width=0.62\linewidth]{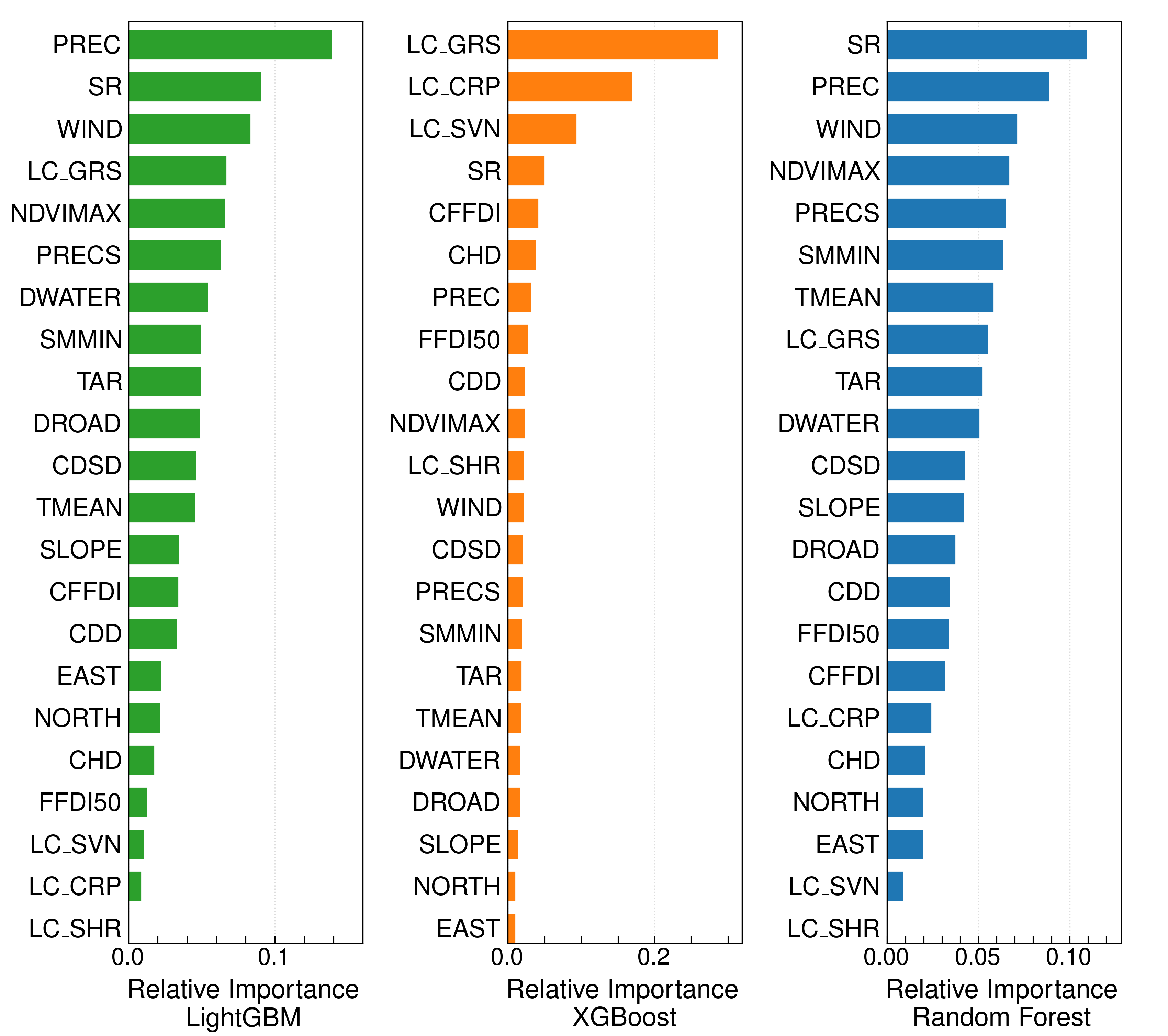}
	\par\vspace{3pt}
	\refstepcounter{figure}\label{fig:physical_feature_importance}
	\begin{minipage}{0.92\linewidth}
		\small\textbf{Fig.~\thefigure:} Relative feature importance of physical variables in tree-based cross-sectional susceptibility models.
	\end{minipage}
\end{center}
\end{samepage}
\endgroup

\clearpage
% Keep Tables S4 and S5 close together and vertically centre them as a group.
\thispagestyle{supplementary}
\begingroup
\setlength{\intextsep}{4pt}
\setlength{\textfloatsep}{4pt}
\null\vfill
\begin{samepage}

\begin{center}
    \refstepcounter{table}\label{tab:transfer_sites}
    \begin{minipage}{\linewidth}
        \normalsize\textbf{Table~\thetable}\par
        Summary of transfer sites used in the transferability analysis, including study area extent and labelled sample counts. A 1:1 positive-to-negative ratio is maintained at each site.
    \end{minipage}
    \par\vspace{3pt}
    \setlength{\tabcolsep}{5pt}
    \begin{tabular}{@{} l r r r @{}}
        \toprule
        \textbf{Transfer Site} & 
        \textbf{Area (km$^2$)} & 
        \textbf{Positive Samples} & 
        \textbf{Total Samples} \\
        \midrule
        ACT Canberra--Namadgi                    & 3{,}172 & 613   & 1{,}226 \\
        NSW Western Sydney--Blue Mountains       & 5{,}002 & 1{,}292 & 2{,}584 \\
        NT Alice Springs--West MacDonnell        & 2{,}560 & 375   & 750   \\
        QLD Beaudesert--Lamington--Mt Lindesay   & 3{,}808 & 305   & 610   \\
        QLD Kuranda--Mareeba--Malanda            & 2{,}914 & 410   & 820   \\
        SA Kangaroo Island                       & 4{,}385 & 792   & 1{,}584 \\
        TAS Forcett--Dunalley                    & 594     & 36    & 72    \\
        WA Leeuwin--Boranup                      & 1{,}829 & 138   & 276   \\
        \bottomrule
    \end{tabular}
\end{center}

\vspace{6pt}
\begin{center}
    \refstepcounter{table}\label{tab:metric_mean_sd}
    \begin{minipage}{\linewidth}
        \normalsize\textbf{Table~\thetable}\par
        Test-set performance metrics (mean $\pm$ standard deviation, \%) across 10 independent train-test splits.
    \end{minipage}
    \par\vspace{3pt}
    \setlength{\tabcolsep}{4pt}
    \begin{tabular}{@{} l l ccccc @{}}
        \toprule
        \multirow{2}{*}{\textbf{Model}} & 
        \multirow{2}{*}{\textbf{Variables}} & 
        \multicolumn{5}{c}{\textbf{Mean $\pm$ SD (\%)}} \\
        \cmidrule(lr){3-7}
        & & \textbf{Accuracy} & \textbf{ROC-AUC} & \textbf{F1} & \textbf{Recall} & \textbf{Specificity} \\
        \midrule
        \multirow{2}{*}{Random Forest}
            & Physical  & 89.41 $\pm$ 0.43 & 95.67 $\pm$ 0.20 & 89.10 $\pm$ 0.46 & 86.57 $\pm$ 0.65 & 92.26 $\pm$ 0.50 \\
            & Embedding & 83.97 $\pm$ 0.26 & 91.94 $\pm$ 0.28 & 83.91 $\pm$ 0.29 & 83.63 $\pm$ 0.72 & 84.31 $\pm$ 0.64 \\
        \addlinespace[2pt]
        \multirow{2}{*}{XGBoost}
            & Physical  & 89.65 $\pm$ 0.39 & 95.80 $\pm$ 0.18 & 89.51 $\pm$ 0.40 & 88.34 $\pm$ 0.61 & 90.95 $\pm$ 0.47 \\
            & Embedding & 85.61 $\pm$ 0.26 & 92.95 $\pm$ 0.20 & 85.44 $\pm$ 0.29 & 84.48 $\pm$ 0.53 & 86.74 $\pm$ 0.31 \\
        \addlinespace[2pt]
        \multirow{2}{*}{LightGBM}
            & Physical  & 90.25 $\pm$ 0.35 & 96.14 $\pm$ 0.19 & 90.05 $\pm$ 0.37 & 88.28 $\pm$ 0.50 & 92.21 $\pm$ 0.36 \\
            & Embedding & 85.71 $\pm$ 0.34 & 93.17 $\pm$ 0.21 & 85.55 $\pm$ 0.38 & 84.59 $\pm$ 0.70 & 86.84 $\pm$ 0.37 \\
        \addlinespace[2pt]
        \multirow{2}{*}{TabPFN}
            & Physical  & 92.32 $\pm$ 0.28 & 97.21 $\pm$ 0.15 & 92.22 $\pm$ 0.29 & 91.07 $\pm$ 0.46 & 93.56 $\pm$ 0.32 \\
            & Embedding & 88.87 $\pm$ 0.25 & 95.49 $\pm$ 0.18 & 88.58 $\pm$ 0.26 & 86.35 $\pm$ 0.42 & 91.38 $\pm$ 0.37 \\
        \addlinespace[2pt]
        \multirow{2}{*}{MLP}
            & Physical  & 85.48 $\pm$ 0.36 & 92.51 $\pm$ 0.32 & 84.89 $\pm$ 0.39 & 81.58 $\pm$ 1.14 & 89.38 $\pm$ 1.21 \\
            & Embedding & 86.40 $\pm$ 0.40 & 93.58 $\pm$ 0.27 & 86.08 $\pm$ 0.44 & 84.11 $\pm$ 1.25 & 88.69 $\pm$ 1.26 \\
        \midrule
        \multirow{2}{*}{CNN9}
            & Physical  & 87.92 $\pm$ 0.41 & 94.48 $\pm$ 0.27 & 87.81 $\pm$ 0.41 & 86.96 $\pm$ 0.78 & 88.89 $\pm$ 0.91 \\
            & Embedding & 88.36 $\pm$ 0.34 & 94.99 $\pm$ 0.23 & 88.08 $\pm$ 0.39 & 86.05 $\pm$ 1.19 & 90.67 $\pm$ 1.14 \\
        \addlinespace[2pt]
        \multirow{2}{*}{CNN17}
            & Physical  & 90.54 $\pm$ 0.64 & 96.00 $\pm$ 0.32 & 90.45 $\pm$ 0.69 & 89.65 $\pm$ 1.26 & 91.44 $\pm$ 0.72 \\
            & Embedding & 90.81 $\pm$ 0.65 & 96.20 $\pm$ 0.28 & 90.74 $\pm$ 0.67 & 89.99 $\pm$ 1.15 & 91.63 $\pm$ 1.19 \\
        \addlinespace[2pt]
        \multirow{2}{*}{CNN25}
            & Physical  & 91.91 $\pm$ 0.58 & 96.72 $\pm$ 0.24 & 91.86 $\pm$ 0.56 & 91.29 $\pm$ 0.80 & 92.52 $\pm$ 1.11 \\
            & Embedding & 91.66 $\pm$ 0.56 & 96.63 $\pm$ 0.29 & 91.62 $\pm$ 0.55 & 91.14 $\pm$ 1.42 & 92.19 $\pm$ 1.63 \\
        \midrule
        \multirow{2}{*}{MLP (temporal)}
            & Physical  & 85.92 $\pm$ 0.35 & 93.06 $\pm$ 0.29 & 85.38 $\pm$ 0.41 & 82.23 $\pm$ 1.04 & 89.62 $\pm$ 0.97 \\
            & Embedding & 87.47 $\pm$ 0.30 & 94.24 $\pm$ 0.26 & 87.15 $\pm$ 0.32 & 85.05 $\pm$ 0.93 & 89.88 $\pm$ 0.96 \\
        \addlinespace[2pt]
        \multirow{2}{*}{CNN9 (temporal)}
            & Physical  & 86.46 $\pm$ 0.51 & 93.48 $\pm$ 0.46 & 86.18 $\pm$ 0.73 & 84.59 $\pm$ 2.83 & 88.32 $\pm$ 2.46 \\
            & Embedding & 88.53 $\pm$ 0.47 & 95.11 $\pm$ 0.24 & 88.43 $\pm$ 0.41 & 87.70 $\pm$ 1.87 & 89.36 $\pm$ 2.34 \\
        \addlinespace[2pt]
        \multirow{2}{*}{CNN17 (temporal)}
            & Physical  & 90.47 $\pm$ 0.61 & 95.87 $\pm$ 0.29 & 90.43 $\pm$ 0.60 & 89.97 $\pm$ 1.02 & 90.97 $\pm$ 1.13 \\
            & Embedding & 89.79 $\pm$ 0.69 & 95.93 $\pm$ 0.37 & 89.84 $\pm$ 0.59 & 90.25 $\pm$ 1.29 & 89.32 $\pm$ 1.18 \\
        \addlinespace[2pt]
        \multirow{2}{*}{CNN25 (temporal)}
            & Physical  & 90.83 $\pm$ 0.73 & 96.05 $\pm$ 0.34 & 90.67 $\pm$ 0.80 & 89.22 $\pm$ 1.60 & 92.44 $\pm$ 1.02 \\
            & Embedding & 91.48 $\pm$ 0.80 & 96.45 $\pm$ 0.30 & 91.43 $\pm$ 0.83 & 90.94 $\pm$ 1.77 & 92.02 $\pm$ 1.69 \\
        \bottomrule
    \end{tabular}
\end{center}

\end{samepage}
\vfill\null
\endgroup

% Fig. S13 occupies a dedicated page and is vertically centred.
\clearpage
\thispagestyle{supplementary}
\null\vfill
\begin{center}
	\includegraphics[width=0.76\linewidth]{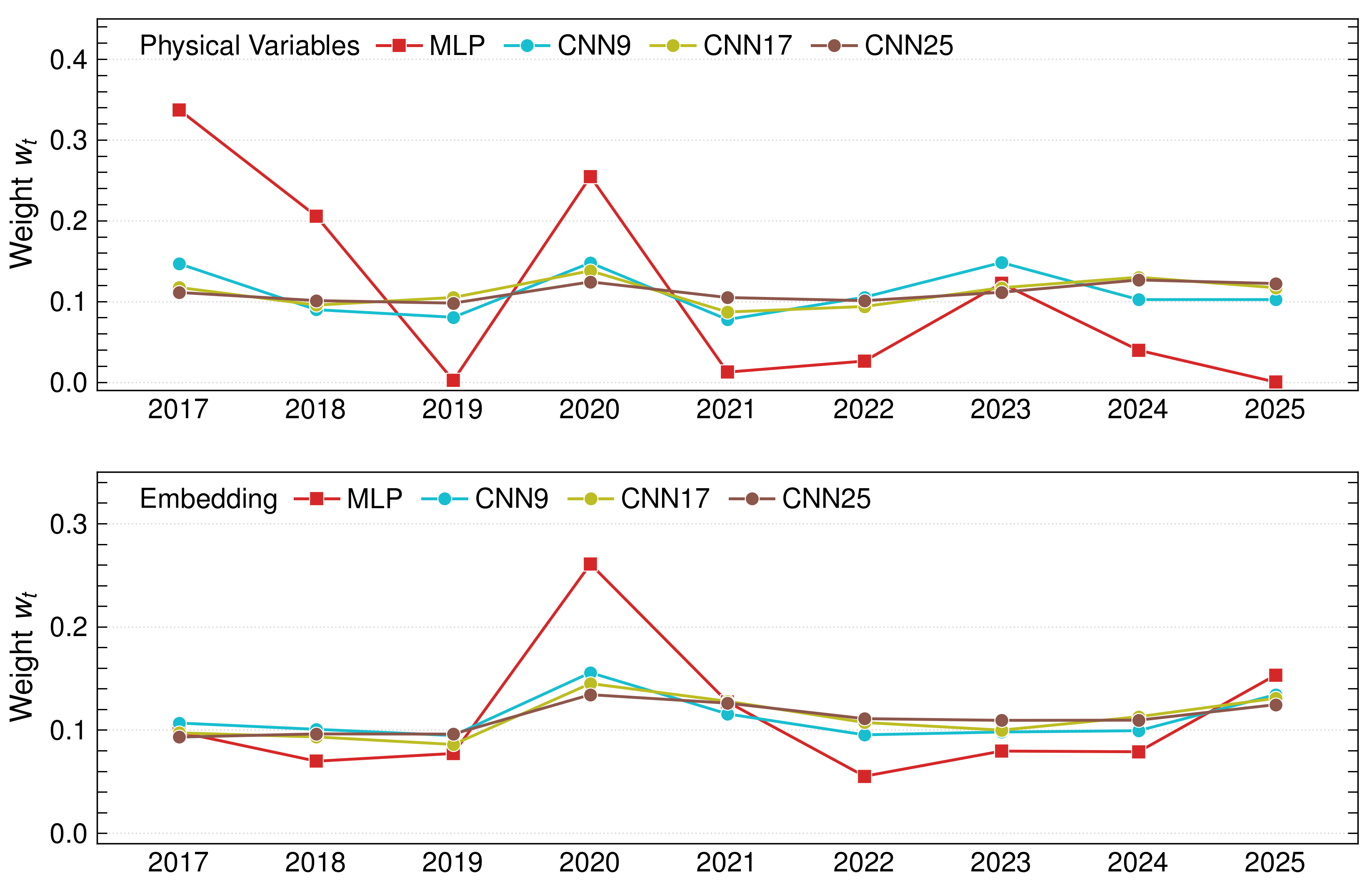}
	\par\vspace{3pt}
	\refstepcounter{figure}\label{fig:temporal_weights}
	\begin{minipage}{0.92\linewidth}
		\small\textbf{Fig.~\thefigure:} Learned yearly weights for sequence-aware physical-variable and embedding-based models (mean across 10 independent runs).
	\end{minipage}
\end{center}
\vfill\null
\clearpage

% Figs. S14 and S15 each occupy a dedicated, vertically centred page.
\thispagestyle{supplementary}
\null\vfill
\begin{center}
	\includegraphics[width=0.96\linewidth]{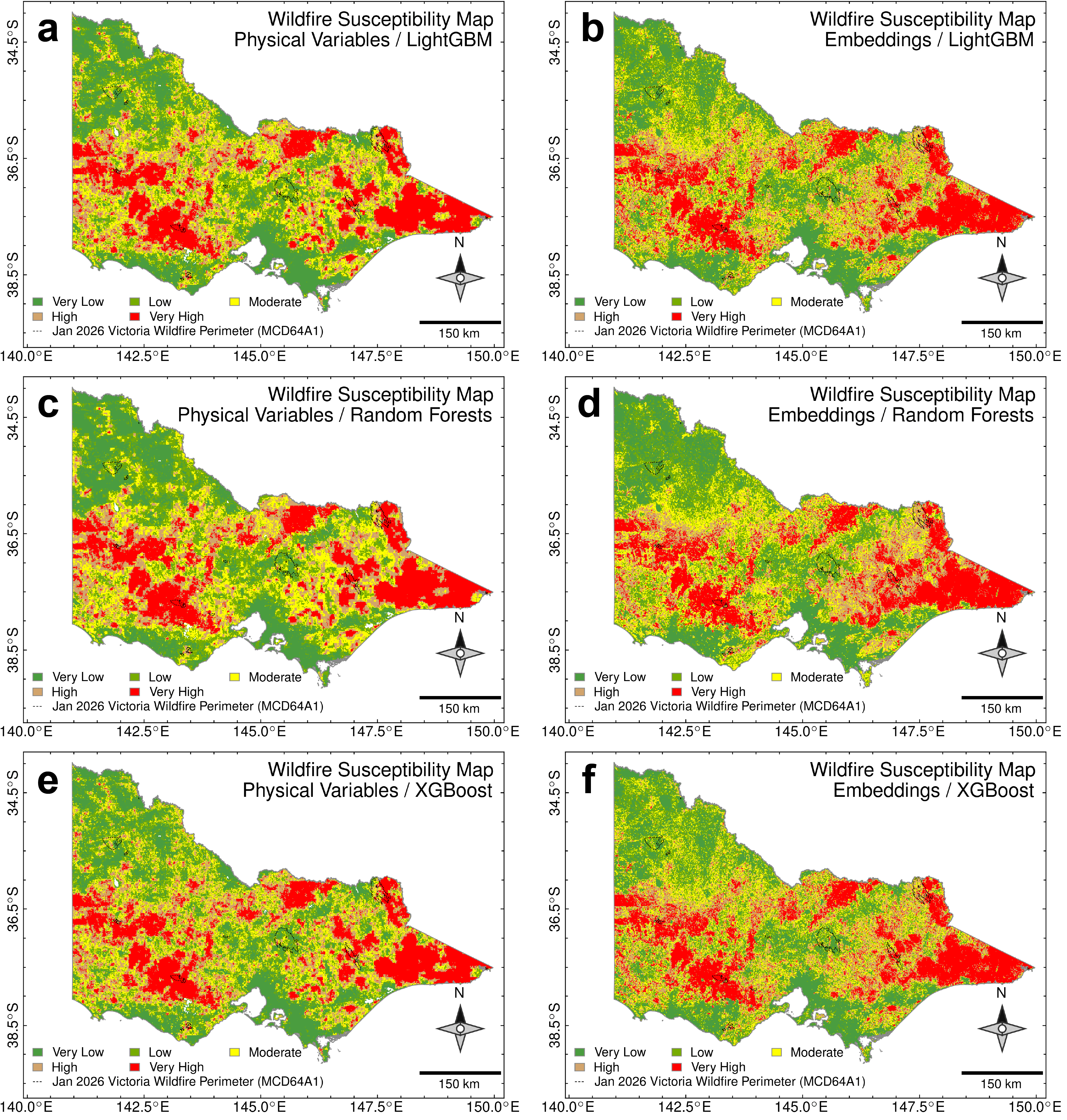}
	\par\vspace{3pt}
	\refstepcounter{figure}\label{fig:composite_trees}
	\begin{minipage}{0.92\linewidth}
		\small\textbf{Fig.~\thefigure:} Supplementary susceptibility maps from tree-based wildfire susceptibility models.
	\end{minipage}
\end{center}
\vfill\null
\clearpage

\thispagestyle{supplementary}
\null\vfill
\begin{center}
	\includegraphics[width=0.96\linewidth]{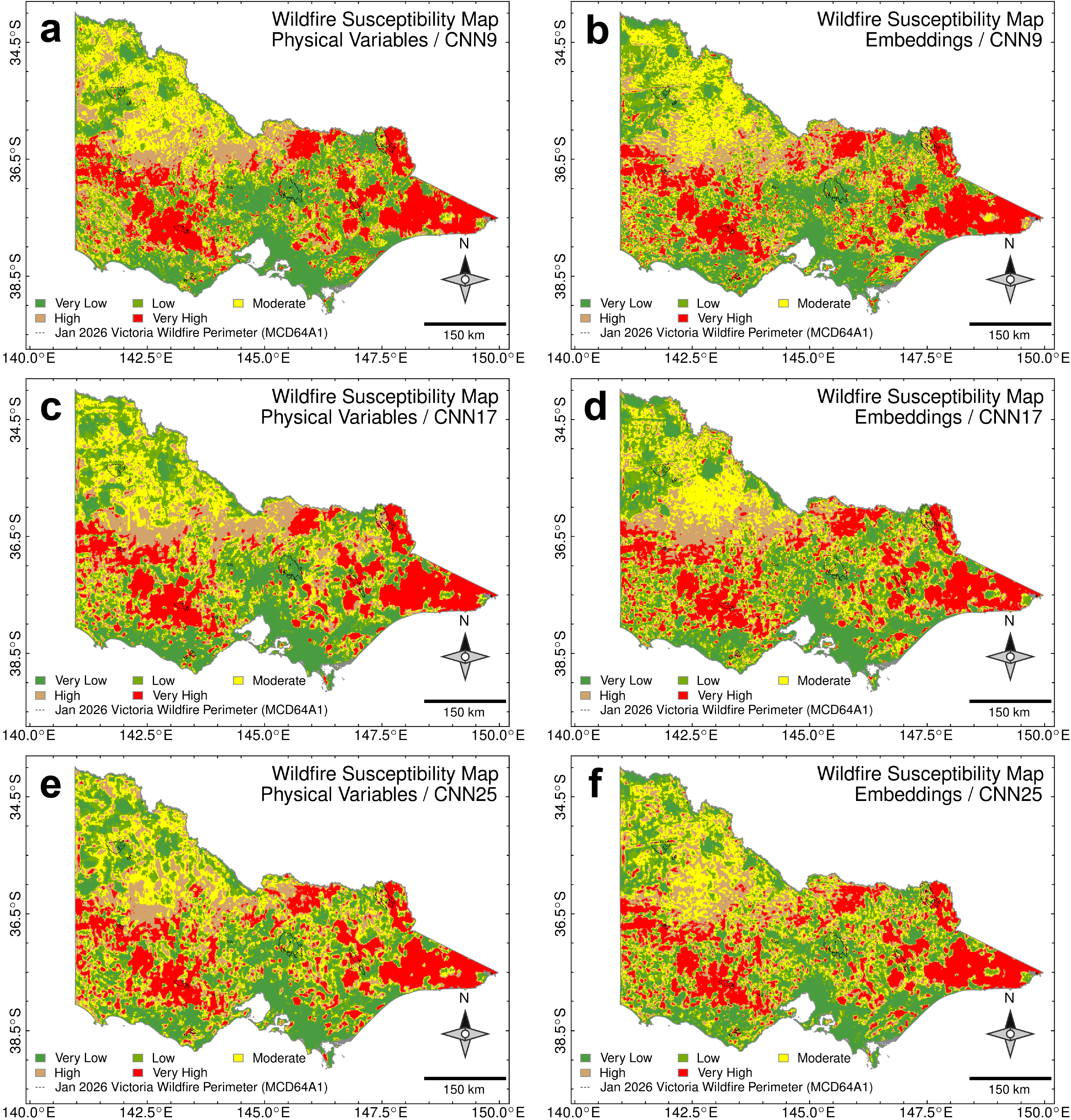}
	\par\vspace{3pt}
	\refstepcounter{figure}\label{fig:composite_cnn}
	\begin{minipage}{0.92\linewidth}
		\small\textbf{Fig.~\thefigure:} Supplementary susceptibility maps from CNN wildfire susceptibility models.
	\end{minipage}
\end{center}
\vfill\null
\clearpage

% Figs. S16 and S17 share one page and are vertically centred as a group.
\thispagestyle{supplementary}
\null\vfill
\begin{samepage}
\begin{center}
	\includegraphics[width=0.67\linewidth]{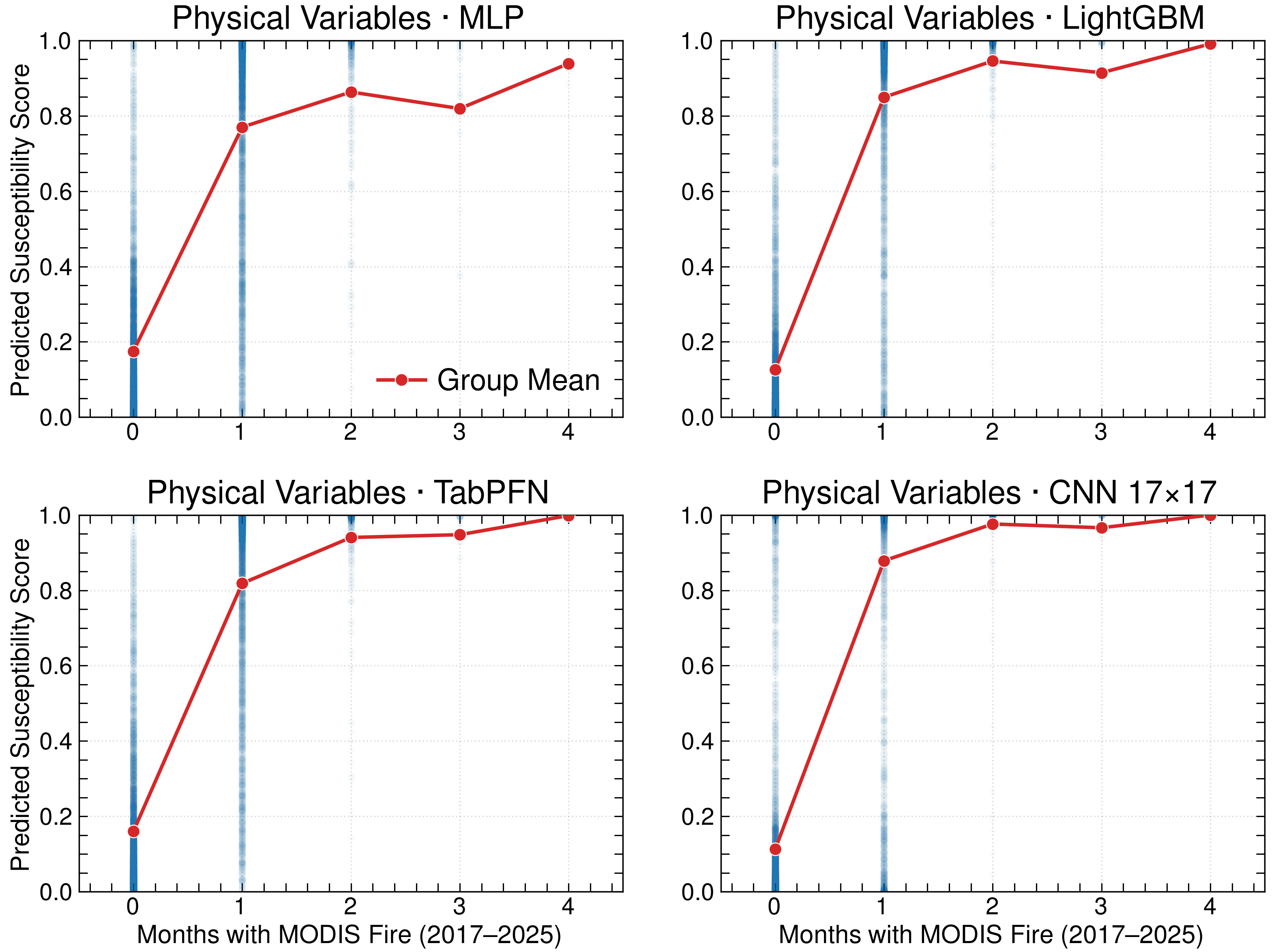}
	\par\vspace{3pt}
	\refstepcounter{figure}\label{fig:scorevsfirecountphysical}
	\begin{minipage}{0.92\linewidth}
		\small\textbf{Fig.~\thefigure:} Relationship between predicted susceptibility scores and the number of historical MODIS fire-occurrence months for physical-variable susceptibility models.
	\end{minipage}
\end{center}

\vspace{-0.5em}
\begin{center}
	\includegraphics[width=0.67\linewidth]{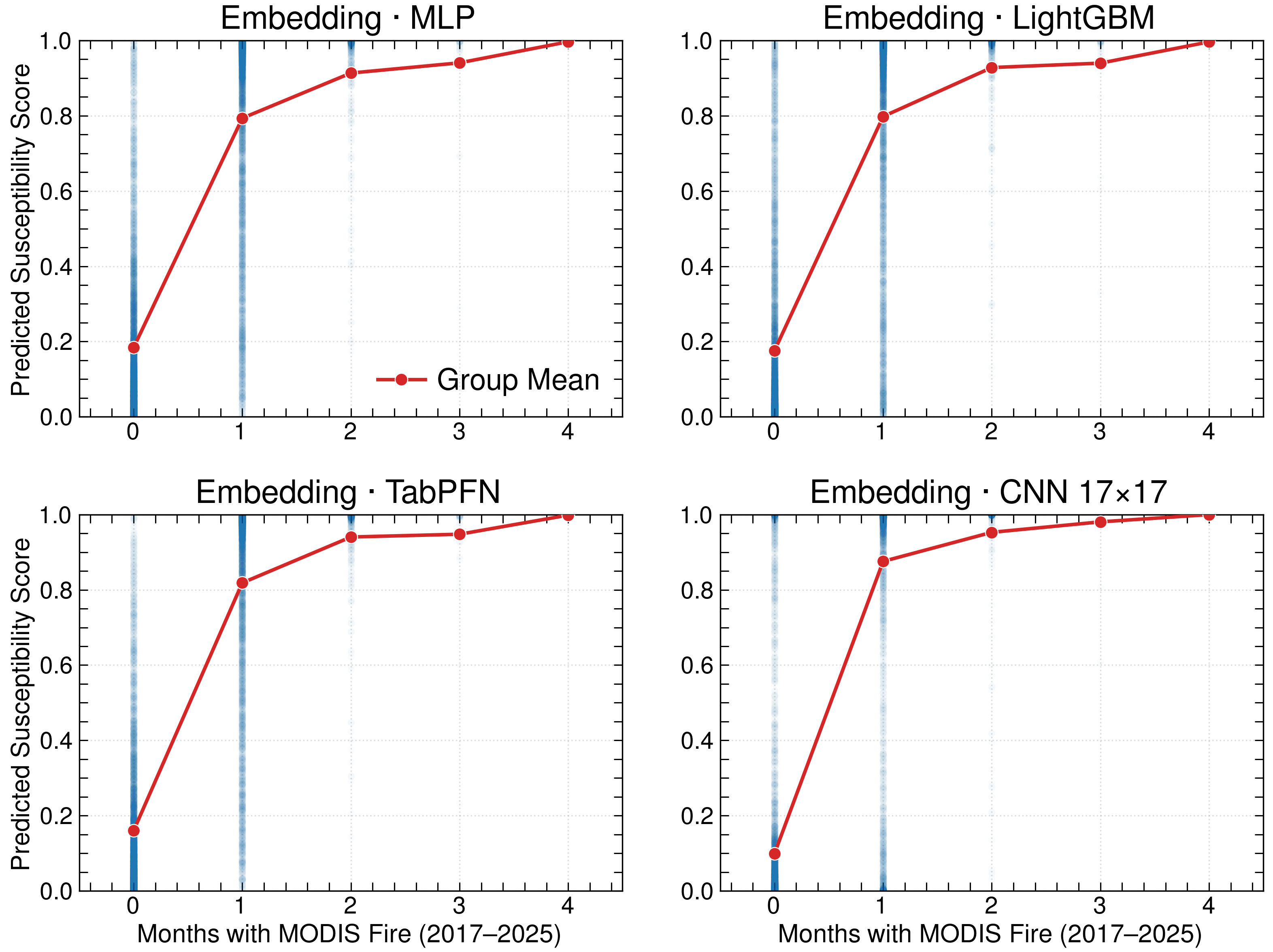}
	\par\vspace{3pt}
	\refstepcounter{figure}\label{fig:scorevsfirecountpanelembedding}
	\begin{minipage}{0.92\linewidth}
		\small\textbf{Fig.~\thefigure:} Relationship between predicted susceptibility scores and the number of historical MODIS fire-occurrence months for embedding-based susceptibility models.
	\end{minipage}
\end{center}
\end{samepage}
\vfill\null
\clearpage

\begin{figure}[htbp]
	\centering
	\includegraphics[width=0.7\linewidth]{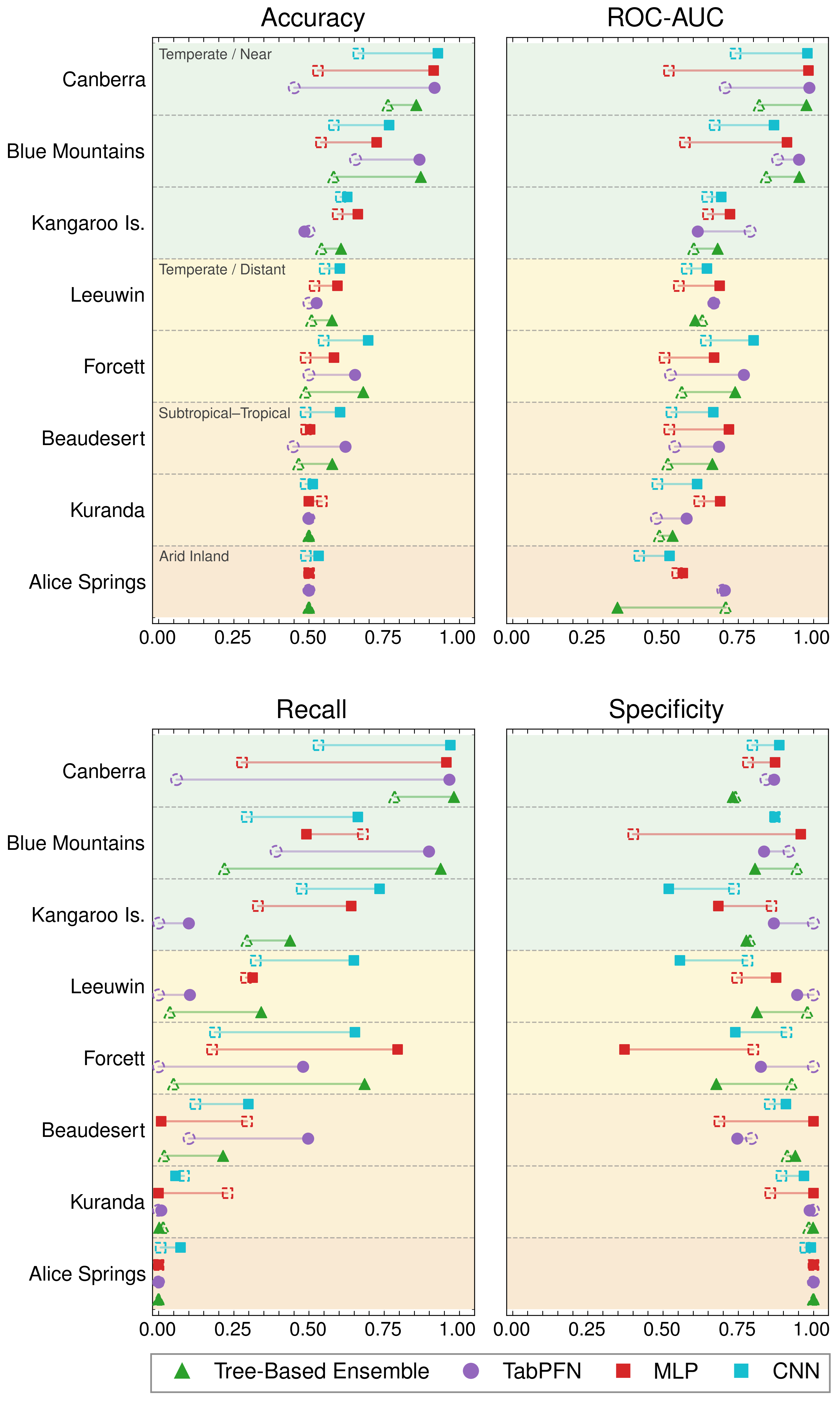}
	\caption{Absolute transfer performance across target regions and model families. For each downstream model family, filled markers show models trained on AEF embeddings and hollow markers show models trained on physical variables.}
	\label{fig:transfer_absolute}
\end{figure}

\clearpage

% Numerical values and run-to-run variability underlying Fig. 7.
\clearpage
\thispagestyle{supplementary}
\begingroup
\null\vfill
\begin{center}
	\refstepcounter{table}\label{tab:transfer_delta_sd}
	\begin{minipage}{\linewidth}
		\normalsize\textbf{Table~\thetable}\par
		Relative transfer performance changes underlying Fig.~\ref{fig:transfer_delta}, calculated using Eq.~\ref{eq:transfer_delta} (mean $\pm$ standard deviation, \%) across 10 independent runs. Values are averaged across model families within each feature representation; negative values indicate performance degradation relative to Victoria.
	\end{minipage}
	\par\vspace{5pt}
	\small
	\setlength{\tabcolsep}{4pt}
	\renewcommand{\arraystretch}{1.03}
	\begin{tabular}{@{} >{\raggedright\arraybackslash}p{4.4cm} l cccc @{}}
		\toprule
		\multirow{2}{*}{\textbf{Transfer Site}} &
		\multirow{2}{*}{\textbf{Variables}} &
		\multicolumn{4}{c}{\textbf{Relative Change, Mean $\pm$ SD (\%)}} \\
		\cmidrule(lr){3-6}
		& & \textbf{Accuracy} & \textbf{ROC-AUC} & \textbf{Recall} & \textbf{Specificity} \\
		\midrule
		\multirow{2}{4.4cm}{Canberra--Namadgi}
		& Physical  & $-32.58 \pm 1.24$ & $-27.04 \pm 2.73$ & $-52.48 \pm 4.04$ & $-13.55 \pm 2.22$ \\
		& Embedding & $  3.18 \pm 0.52$ & $  3.98 \pm 0.27$ & $ 12.86 \pm 0.79$ & $ -6.14 \pm 1.02$ \\
		\addlinespace[2pt]
		\multirow{2}{4.4cm}{Western Sydney--Blue Mountains}
		& Physical  & $-34.12 \pm 3.96$ & $-22.35 \pm 5.90$ & $-54.02 \pm 5.13$ & $-14.56 \pm 5.60$ \\
		& Embedding & $ -7.78 \pm 2.14$ & $ -2.35 \pm 1.03$ & $-12.91 \pm 5.58$ & $ -2.91 \pm 1.72$ \\
		\addlinespace[2pt]
		\multirow{2}{4.4cm}{Kangaroo Island}
		& Physical  & $-37.16 \pm 3.38$ & $-29.56 \pm 3.24$ & $-68.31 \pm 11.34$ & $ -7.50 \pm 4.40$ \\
		& Embedding & $-32.04 \pm 1.17$ & $-28.15 \pm 1.50$ & $-44.43 \pm 4.21$ & $-20.16 \pm 3.93$ \\
		\addlinespace[2pt]
		\multirow{2}{4.4cm}{Leeuwin--Boranup}
		& Physical  & $-41.82 \pm 1.01$ & $-36.32 \pm 1.60$ & $-81.01 \pm 5.49$ & $ -4.28 \pm 4.72$ \\
		& Embedding & $-34.41 \pm 1.42$ & $-30.90 \pm 1.44$ & $-59.32 \pm 3.12$ & $-10.56 \pm 2.20$ \\
		\addlinespace[2pt]
		\multirow{2}{4.4cm}{Forcett--Dunalley}
		& Physical  & $-43.32 \pm 2.19$ & $-41.42 \pm 3.69$ & $-87.86 \pm 6.31$ & $ -0.63 \pm 3.27$ \\
		& Embedding & $-25.45 \pm 2.32$ & $-21.12 \pm 1.83$ & $-23.79 \pm 8.05$ & $-27.03 \pm 3.93$ \\
		\addlinespace[2pt]
		\multirow{2}{4.4cm}{Beaudesert--Lamington--Mt Lindesay}
		& Physical  & $-47.00 \pm 2.39$ & $-44.83 \pm 2.70$ & $-84.25 \pm 4.67$ & $-11.21 \pm 2.39$ \\
		& Embedding & $-34.24 \pm 1.04$ & $-27.55 \pm 1.50$ & $-70.61 \pm 2.97$ & $  0.79 \pm 2.24$ \\
		\addlinespace[2pt]
		\multirow{2}{4.4cm}{Kuranda--Mareeba--Malanda}
		& Physical  & $-43.10 \pm 0.99$ & $-45.66 \pm 2.05$ & $-90.17 \pm 4.86$ & $  2.04 \pm 3.79$ \\
		& Embedding & $-42.63 \pm 0.49$ & $-36.07 \pm 1.97$ & $-98.14 \pm 0.95$ & $ 10.73 \pm 0.49$ \\
		\addlinespace[2pt]
		\multirow{2}{4.4cm}{Alice Springs--West MacDonnell}
		& Physical  & $-44.33 \pm 0.35$ & $-37.77 \pm 4.63$ & $-99.81 \pm 0.29$ & $  8.66 \pm 1.17$ \\
		& Embedding & $-42.02 \pm 1.32$ & $-43.39 \pm 7.04$ & $-97.94 \pm 2.87$ & $ 11.74 \pm 0.51$ \\
		\bottomrule
	\end{tabular}
\end{center}
\vfill\null
\endgroup
\clearpage

\thispagestyle{supplementary}
\putbib[references]
\clearpage
\end{bibunit}

% Save the independent Supplementary Materials page total for the next run.
\xdef\SupplementLastPage{\number\numexpr\value{page}-1\relax}
\makeatletter
\immediate\write\@auxout{%
  \string\gdef\string\SupplementLastPage{\SupplementLastPage}}
\makeatother

\endgroup

\end{document}